\documentclass[a4paper,preprint,aps,prd,superscriptaddress,tightenlines]{revtex4-1}
\usepackage[latin9]{inputenc}
\usepackage{color}
\usepackage{amsmath}
\usepackage{amssymb}
\usepackage{graphicx}

\makeatletter

\pdfpageheight\paperheight
\pdfpagewidth\paperwidth

\providecommand{\tabularnewline}{\\}

\usepackage{hyperref}
 
\hypersetup{
   citecolor=blue,
   linkcolor=blue,
   urlcolor=blue
 }

\makeatother

\begin{document}
\title{Constraining light dark matter upscattered by ultrahigh-energy cosmic
rays }
\author{Chen Xia}
\affiliation{CAS key laboratory of theoretical Physics, Institute of Theoretical
Physics, Chinese Academy of Sciences, Beijing 100190, China. University
of Chinese Academy of Sciences, Beijing, 100190, China.}
\author{Yan-Hao Xu}
\affiliation{CAS key laboratory of theoretical Physics, Institute of Theoretical
Physics, Chinese Academy of Sciences, Beijing 100190, China. University
of Chinese Academy of Sciences, Beijing, 100190, China.}
\author{Yu-Feng Zhou}
\affiliation{CAS key laboratory of theoretical Physics, Institute of Theoretical
Physics, Chinese Academy of Sciences, Beijing 100190, China. University
of Chinese Academy of Sciences, Beijing, 100190, China.}
\affiliation{School of Fundamental Physics and Mathematical Sciences, Hangzhou
Institute for Advanced Study, UCAS, Hangzhou 310024, China. }
\affiliation{International Centre for Theoretical Physics Asia-Pacific, Beijing/Hangzhou,
China.}
\begin{abstract}
Light halo dark matter (DM) particles upscattered by high-energy cosmic
rays (CRs) can be energetic, and become detectable by conventional
direct detection experiments. The current constraints derived from
space-based direct CR measurements can reach $\mathcal{O}(10^{-31})\text{ cm}^{2}$
for a constant DM-nucleon scattering cross section. We show that if
the CR energy spectrum follows a power law of type $\sim E^{-3}$,
the derived constraints on the scattering cross section will be highly
insensitive to DM particle mass. This suggests that ultrahigh-energy
CRs (UHECRs) indirectly measured by ground-based detectors can be
used to place constraints on ultralight DM particles, as $E^{-3}$
is a very good approximation of the UHECR energy spectrum up to energy
$\sim10^{20}\text{ eV}$.  Using the recent UHECR flux data, we
show that the current constraints derived from space-based CR measurements
can in principle be extended to ultralight DM particles far below
eV scale . 
\end{abstract}
\date{\today}
\maketitle

\section{Introduction }

Although compelling astrophysical evidence supports the existence
of dark matter (DM) in the Universe, whether or not DM participates
non-gravitational interactions is still an open question. The majority
of the current DM direct detection (DD) experiments search for nuclear
recoil signals from the scatterings between the halo DM particle and
target nucleus.  As the typical detection threshold of the current
experiments is of $\text{\ensuremath{\mathcal{O}(}keV)}$, searching
for light halo DM below GeV is in general challenging. The reason
is that for lighter halo DM particles the kinetic energy is lower,
and the energy transferred to the target nuclei is suppressed. For
instance, for a DM particle with mass $m_{\chi}\sim1$ GeV and a typical
DM escape velocity $\sim540\ \text{km s}^{-1}$, the elastic scattering
off a target nucleus with mass $\sim100$ GeV leads to a maximal recoil
energy $\sim0.06$\,keV which is significantly lower than the typical
detection threshold. Several physical processes have been considered
to lower the detection threshold such as using additional photon emission
in the inelastic scattering process \citep{Kouvaris:2016afs} and
the Migdal effect \citep{Ibe:2017yqa,Dolan:2017xbu}, etc.. The same
DM-nucleus scattering process may leave imprints in some cosmological
and astrophysical observables, which can be used to place constraints
on the scattering cross section. The resulting constraints are in
general much weaker but can be applied to lower DM particle masses
unreachable to the DD experiments. For instance, from the spectral
distortion of the cosmic microwave background (CMB), a constraint
of $\sigma_{\chi p}\lesssim5\times10^{-27}\text{cm}^{2}$ for $m_{\chi}$
at 1~keV-TeV can be obtained \citep{Gluscevic:2017ywp}; the constraints
from the population of Milky Way (MW) satellite galaxies can reach
$\sigma_{\chi p}\lesssim6\times10^{-30}\text{cm}^{2}$ for $m_{\chi}\lesssim10\text{ keV}$
\citep{Nadler:2019zrb}; and the measurement of the gas cooling rate
of the Leo T dwarf galaxy can also lead to a constraint of $\sigma_{\chi p}\lesssim3\times10^{-25}\text{cm}^{2}$
for $m_{\chi}\lesssim$ 1~MeV \citep{Wadekar:2019xnf,Bhoonah:2018wmw}.\emph{ }

Recently, it was shown that important constraints can be derived from
the scattering between cosmic ray (CR) particles and DM particles
in the local Universe. High-energy CR particles in the Galaxy can
scatter off halo DM particles, which results in the energy-loss of
CRs \citep{Cappiello:2018hsu}, the production of $\gamma$-rays \citep{Cyburt:2002uw,Hooper:2018bfw}
and energy-boost of DM particles \citep{Bringmann:2018cvk,Ema:2018bih,Cappiello:2019qsw},
etc.. In the last process, a small but irreducible component of DM
(referred to as CRDM) can obtain very high kinetic energies. These
energetic CRDM particles can scatter again with the target nuclei
in underground detectors, and deposit sufficient energy to cross the
detection threshold, which greatly extend the sensitivity of the current
DD experiments to sub-GeV DM particles \citep{Bringmann:2018cvk,Ema:2018bih,Dent:2019krz,Bondarenko:2019vrb,Cappiello:2019qsw,Ge:2020yuf,Zhang:2020htl,Wang:2019jtk}.
It has been shown that in this approach the constraints on constant
DM-nucleon (DM-electron) spin-independent scattering cross section
can reach $\sim10^{-31}(10^{-34})\,\text{cm}^{2}$ for DM particle
mass down to at least $\sim0.1$ MeV ($\sim1$ eV) \citep{Bringmann:2018cvk,Ema:2018bih}
(for constraints on energy-dependent cross sections, see e.g. \citep{Dent:2019krz,Bondarenko:2019vrb}).

It is of interest to explore the potential of this approach in constraining
even lighter DM particles far below MeV scale, as some well-motivated
DM candidates such as QCD axions and axion-like particles can be extremely
light. Constraining lighter DM requires better information on the
CR spectra at higher energies. Note, however, that all the current
analysis \citep{Bringmann:2018cvk,Ema:2018bih,Dent:2019krz,Bondarenko:2019vrb,Cappiello:2019qsw,Ge:2020yuf,Zhang:2020htl,Wang:2019jtk}
on the detection of CRDM adopted the CR fluxes from either the parametrizations
in\,\citep{DellaTorre:2016jjf,Bisschoff:2019lne} or the GALPROP
code\, \citep{Strong:1998pw,Moskalenko:1997gh}, which are inferred
from the space-based direct CR measurements (e.g. PAMELA\,\citep{Adriani:2011cu,Adriani:2013as},
AMS-02 \citep{Aguilar:2015ooa,Aguilar:2015ctt} and CREAM-I \citep{Yoon:2011aa}
etc.). For current space-based experiments the CR fluxes which can
be measured with reasonable precision are typically with energy $\lesssim200$\,TeV
(see also the data of CREAM-III \citep{Yoon:2017qjx}, CALET \citep{Adriani:2019aft}
and DAMPE \citep{An:2019wcw}). Towards higher energies, the statistic
uncertainties increase rapidly due to the limited acceptance of space-based
experiments\,\citep{Nilsen:1997mv,Derbina:2005ta,Atkin:2018wsp}.
 Naively extrapolating these analyses to higher energies will leads
to incorrect conclusions, as the spectral feature of the CR flux start
to change above 1 PeV. Alternatively, high-energy CR can be measured
indirectly by ground-based air-shower detectors. Despite larger uncertainties
in energy scale and mass resolution, this approach can measure the
CR flux to much higher energy due to the huge acceptances.  For detecting
lighter DM particles, the local DM number density is higher. However,
the energy transfer from the scattering process become less efficient,
and the CR flux is known to decrease rapidly towards higher energies.
Whether or not ultrahigh-energy CRs (UHECRs, defined as CR with total
energy $E>$ PeV) can be used to place useful constraints on ultralight
DM will depend strongly on the spectral feature of the UHECR flux.

In this work, we show that as long as the energy spectrum of CR flux
follows a power law $\sim E^{-\alpha}$ with $\alpha\lesssim3$, the
derived constraints on the DM-nucleon scattering cross section will
not decrease towards lower DM mass. In the limit of $\alpha=3$, the
constraints will DM mass independent. This justifies using UHECR
to place stringent constraints on ultralight DM particles, as the
UHECR all-particle spectrum above the ``knee'' structure (at $\sim3$
PeV) can be well-approximated by a power law with $\alpha\approx3$.
From the recent UHECR nucleus flux data, we obtain the following results:
the constraints on the spin-independent DM-nucleon scattering cross
section can be $\lesssim10^{-(32-31)}\ \text{cm}^{2}$ for DM particle
mass down to extremely small value $\sim10^{-12}\text{ eV}$, which
expands the currently known constraints derived from space-based direct
CR measurements by around ten orders of magnitude in DM mass, and
close a large previously unconstrained parameter space;\emph{ }the
most stringent constraints are found to be at DM mass $\sim10^{-5}$
eV and $\sim10^{-11}$ eV, due to the ``knee'' and ``toe'' structure
in the UHECR flux, respectively; this CRDM approach will completely
loss sensitivity for DM mass below $10^{-14}$ eV as the UHECR flux
is highly suppressed above $\sim10^{20}$ eV, a phenomena possibly
related to the scatterings between UHECRs and cosmic microwave background
(CMB) photons \citep{Greisen:1966jv,Zatsepin:1966xyz}.  The constraints
obtained in this work are highly model-independent and conservative,
as only the elastic scattering process is required and very conservative
choices of parameters are adpoted. The constraints obtained in this
work are derived based on the observables of the present-day local
Universe, which are complementary to other constraints derived from
the data of earlier epochs of the Universe. 

This paper is organized as follows: In section \ref{sec:CR-upscattered},
we discuss the spectral feature of the DM flux upscattered by UHECRs.
In section \ref{sec:Earth-atten}, we discuss the effect of earth
attenuation of the CRDM kinetic energy. The nuclear recoil sepectrum
and the constraints from direct detection experiment Xenon-1T is discussed
in section \ref{sec:Direct-detection}. We summarize the work and
give some remarks in section \ref{sec:Conclusions}.

\section{CR-upscattered dark matter flux\label{sec:CR-upscattered} }

\subsection{Single CR component case}

In the generic process of elastic scattering between an incident particle
$A$ with kinetic energy $T_{A}$ and a target particle $B$ at rest,
the recoil energy of particle $B$ in the laboratory frame is given
by $T_{B}=T_{B}^{\text{max}}(1-\cos\theta)/2$, where $\theta$ is
the scattering angle of particle $B$ in the center-of-mass (CM) frame.
The maximal recoil energy of particle $B$ is given by
\begin{equation}
\frac{T_{B}^{\text{max}}}{T_{A}}=\left[1+\frac{(m_{B}-m_{A})^{2}}{2m_{B}(T_{A}+2m_{A})}\right]^{-1},\label{eq:Tmax}
\end{equation}
where $m_{A(B)}$ is the mass of particle $A(B)$. We assume that
the scattering is isotropic in the CM frame, such that the differential
cross section $d\sigma_{AB}/dT_{B}$ in the laboratory frame is simply
related to the total cross section $\sigma_{AB}$ as $d\sigma_{AB}/dT_{B}=\sigma_{AB}/T_{B}^{\text{max}}$.
In the case of CR-DM scattering, if the CR particle $i\,(i=\text{H},\text{He},\dots)$
is highly relativistic, i.e., the Lorentz factor $\gamma_{i}\approx T_{i}/m_{i}\gg1$,
but $m_{\chi}$ is small enough such that $\gamma_{i}\ll m_{i}/2m_{\chi}$,
the maximal recoil energy of the CRDM can be approximated as $T_{\chi}^{\text{max}}\approx2m_{\chi}\gamma_{i}^{2}\thinspace$.
The CRDM particle with kinetic energy $T_{\chi}$ can scatter again
with the nucleus $N$ (with mass $m_{N}$) in either the outer crust
of Earth or the detector of the underground DM direct detection experiments.
The maximal recoil energy $T_{N}^{\text{max}}$ of the nucleus which
is also the maximal energy-loss of CRDM particle can be well approximated
as $T_{N}^{\text{max}}\approx2T_{\chi}^{2}/m_{N}$. Note that $T_{N}^{\text{max}}$
is independent of DM particle mass.  

After being upscattered, the CRDM particles travel through the Galaxy
in straight lines as they are not deflected by the interstellar magnetic
fields. The observed flux (number of particles per unit area, time
and solid angle, $dN/dAdtd\Omega$) of CRDM at the surface of Earth
can be approximately written as 
\begin{align}
\frac{d\text{\ensuremath{\Phi}}_{\chi}}{dT_{\chi}} & \approx\frac{\rho_{\chi}^{\text{loc}}\sigma_{\chi i}D_{\text{eff}}F^{2}(Q_{\chi}^{2})}{m_{\chi}}\int_{\gamma_{i}^{\text{min}}(T_{\chi})}^{\infty}\frac{d\gamma_{i}}{T_{\chi}^{\text{max}}}\frac{d\Phi_{i}^{\text{LIS}}}{d\gamma_{i}},\label{eq:CRDM-flux}
\end{align}
where $d\Phi_{i}^{\text{LIS}}/d\gamma_{i}$ is the local interstellar
CR flux measured at Earth. The integration lower limit $\gamma_{i}^{\text{min}}\approx(T_{\chi}/2m_{\chi})^{1/2}$
is the minimal Lorentz factor required to produce $T_{\chi}$. The
form factor $F(Q_{\chi}^{2})$ is evaluated at the momentum transfer
$Q_{\chi}^{2}=2m_{\chi}T_{\chi}$. For very light DM, $F(Q_{\chi}^{2})\approx1$
is an excellent approximation. In the above expression we have assumed
that the CR energy spectrum in the Galactic halo is not significantly
different from that in the local interstellar (LIS) region, i.e.,
$d\Phi_{i}(r)/d\gamma_{i}\approx d\Phi_{i}^{\text{LIS}}/d\gamma_{i}$.
In this case, the information of halo DM density distribution can
be parameterized into a single parameter $D_{\text{eff}}$
\begin{align}
D_{\text{eff}} & \equiv\frac{1}{4\pi\rho_{\chi}^{\text{loc}}}\int_{\text{l.o.s}}\rho_{\chi}dsd\Omega,
\end{align}
 where $\rho_{\chi}^{\text{loc}}\approx0.3\text{ GeV/cm}^{3}$ is
the local DM density in the Solar system, and the integration is performed
along the line-of-sight (l.o.s). For typical DM profiles, the value
of $D_{\text{eff}}$ is around a few kpc. In this work, we make a
conservative choice of $D_{\text{eff}}=1\thinspace\text{kpc}$ as
a benchmark value. 

Let us start with a simple case where the flux of a CR species $i$
can be parametrized by a single power law with index $\alpha_{i}$
and a cutoff at a Lorentz factor $\gamma_{i,\text{cut}}$ as follows
\begin{equation}
\frac{d\Phi_{i}^{\text{LIS}}}{d\gamma_{i}}=\Phi_{i}^{0}\gamma_{i}^{-\alpha_{i}}\exp\left(-\frac{\gamma_{i}}{\gamma_{i,\text{cut }}}\right),\label{eq:CR-flux}
\end{equation}
where $\Phi_{i}^{0}$ is a normalization factor. The power-law behavior
is expected if CRs are accelerated by the diffusive shock waves of
the Galactic supernova-remnants (SNRs) and the pulsar wind, etc.,
and the cutoff represents the maximal energy that can be achieved
by the acceleration process. If $m_{\chi}$ is sufficiently small
such that $\gamma_{i,\text{cut}}^{\text{ }}\ll m_{i}/2m_{\chi}$,
which is easily justified for sub-eV CRDM, the approximation of $T_{\chi}^{\text{max}}\approx2m_{\chi}\gamma_{i}^{2}\thinspace$
can be used in the whole integration range of Eq. (\ref{eq:CRDM-flux}),
and the corresponding CRDM flux can be obtained analytically as follows
\begin{align}
\frac{d\Phi_{\chi}}{dT_{\chi}} & =\frac{\sigma_{\chi i}\rho_{\chi}^{\text{loc}}D_{\text{eff}}\Phi_{i}^{0}F^{2}}{2m_{\chi}^{2}\gamma_{i,\text{cut}}^{\alpha_{i}+1}}\Gamma(-(\alpha_{i}+1),t),\label{eq:DM-flux-general}
\end{align}
where $\Gamma$ is the incomplete $\Gamma$-function, $t=(T_{\chi}/T_{\chi,\text{cut}}^{\text{max}})^{1/2}$
with $T_{\chi,\text{cut}}^{\text{max}}=2m_{\chi}\gamma_{i\text{,cut}}^{2}$
the maximal energy of CRDM upscattered by UHECR at the cutoff $\gamma_{i\text{,cut}}$. 

In the region far below the cutoff, i.e., $T_{\chi}\ll T_{\chi,\text{cut}}^{\text{max}},$
which corresponds to the case where the CR flux is essentially a single
power law $\sim T_{i}^{-\alpha_{i}}$. Using the asymptotic behavior
of the incomplete $\Gamma$-function $\Gamma(a,z)\to-z^{a}/a$ for
$z\ll1$, the CRDM flux can be approximated as
\begin{equation}
\frac{d\Phi_{\chi}}{dT_{\chi}}\approx\frac{2\sigma_{\chi i}\rho_{\chi}^{\text{loc}}D_{\text{eff}}\Phi_{i}^{0}F^{2}}{\alpha_{i}+1}T_{\chi}^{-2}\left(\frac{T_{\chi}}{2m_{\chi}}\right)^{(3-\alpha_{i})/2}\ .\label{eq:CRDM-flux-powerlaw}
\end{equation}
As $F(Q_{\chi}^{2})\approx1$ is a very good approximation for ultralight
DM, the above expression shows that the CRDM flux follows a power
law $\sim T_{\chi}^{-(1+\alpha_{i})/2}$. The mass dependence of the
CRDM flux is proportional to $m_{\chi}^{(\alpha_{i}-3)/2},$ which
shows that as long as the CR flux is hard enough, namely, $\alpha_{i}\lesssim3$,
the resulting CRDM flux will not decrease with decreasing DM mass.\textcolor{black}{{} }

\textcolor{black}{Note that in a wide energy range the CR flux is
close to the case of $\alpha_{i}\approx3$. }\textcolor{black}{Direct
and indirect measurements show that from a few GeV up to the ``knee''
(at $\sim3$ PeV), the primary CR all-particle spectrum approximately
follows a single power law with index $\alpha_{i}\approx2.7$. Above
the ``knee'' the spectrum softens to $\alpha_{i}\approx3.1$. Before
reaching the highest observed energy $\sim10^{20}$ eV, there are
several minor spectral structures such as the ``second knee'' at
$\sim10^{17}$ eV, the ``ankle'' at $\sim8\times10^{18}$ eV and
the ``toe'' at $\sim3\times10^{19}$ eV. The corresponding power-law
indices vary around the $\alpha_{i}\approx3$ case.} \textcolor{black}{Consequently,
the DM upscattered by UHECR in this ultrahigh-energy region should
fluctuate around the power law $T_{\chi}^{-2}$, and is highly insensitive
to DM mass; and the recoil event rate and the derived bounds on $\sigma_{\chi i}$
should be independent of $m_{\chi}$ as well, as the recoil energy
or the energy loss in the $\chi N$ scattering is almost independent
of DM mass. Eq.~(\ref{eq:CRDM-flux-powerlaw}) also suggests that
the CR electrons are less efficient in constraining ultralight DM
particles, as the CR electron flux follows a power law with power
index $\alpha_{e}\approx4$ after $\sim0.9$~TeV~\citep{Aharonian:2009ah,Adriani:2017efm,Ambrosi:2017wek}.}

In a different region where $T_{\chi}$ is close to the cutoff $T_{\chi,\text{cut}}^{\text{max}}$,
the effect of cutoff in CR flux will be significant. Using the asymptotic
behavior of $\Gamma(a,z)\to z^{a-1}e^{-z}$ for large $z$, the CRDM
flux is given by
\begin{equation}
\frac{d\Phi_{\chi}}{dT_{\chi}}\approx\frac{\sigma_{\chi i}\rho_{\chi}^{\text{loc}}D_{\text{eff}}\Phi_{i}^{0}F^{2}}{2m_{\chi}^{2}\gamma_{i,\text{cut}}^{\alpha_{i}+1}}\left(\text{\ensuremath{\frac{T_{\chi}}{T_{\chi,\text{cut}}^{\text{max}}}}}\right)^{-\frac{\alpha_{i}+2}{2}}e^{-\left(\frac{T_{\chi}}{T_{\chi,\text{cut}}^{\text{max}}}\right)^{1/2}}.\label{eq:CRDM_cutoff}
\end{equation}
Since $T_{\chi,\text{cut}}^{\text{max}}$ is proportional to $m_{\chi}$,
lighter CRDM particles will have an earlier cutoff. A final cutoff
in the CR flux around $\sim10^{20}\,\text{eV}$ is expected from the
inelastic scattering between UHECR particles and CMB photons as predicted
by Greisen, Zatsepin and Kuzmin \citep{Greisen:1966jv,Zatsepin:1966xyz},
which is supported by recent observations \citep{Abbasi:2007sv,Castellina:2019huz,Deligny:2020gzq}.
The cutoff in the UHECR flux essentially sets the scale of the minimal
$m_{\chi}$ that can be constrained by this approach.   

\subsection{Multiple CR component case}

The primary CR flux in the ultrahigh-energy region may receive contributions
from different sources such as SNRs, pulsar winds and active galactic
nuclei (AGN), etc. (for recent reviews see e.g. \citep{Kachelriess:2019oqu,Watson:2013cla,Nagano:2000ve}).
The multi-source description is also essential to reproduce the observed
various spectral structures of UHECRs. Thus a realistic description
of the UHECR flux necessarily contains multiple components, which
can be written as $\Phi_{i}=\sum_{j}\Phi_{ij}$ with $j=1,\dots,n$.
For each component $j$, the flux $\Phi_{ij}$ takes the form of Eq.\,(\ref{eq:CR-flux})
with the power index $\alpha_{i}$ and cutoff $\gamma_{i,\text{cut}}$
replaced by $\alpha_{ij}$ and $\gamma_{ij,\text{cut}}$, respectively.
Thus we adopt the following form of the primary CR flux \citep{Gaisser:2013bla}
\begin{equation}
\frac{d\Phi_{i}^{\text{LIS}}}{d\gamma_{i}}=\sum_{j=1}^{n}\Phi_{ij}^{0}\gamma_{i}^{-\alpha_{ij}}\exp\left[-\frac{\gamma_{i}}{\gamma_{ij,\text{cut}}}\right],\label{eq:CR-param-Gaisser}
\end{equation}
where $\Phi_{ij}^{0}$ and $\alpha_{ij}$ are the normalization factors
and power indices, respectively, for a CR species $i$ in the component
$j$.  Following the reasoning of Peters \citep{Peters:1966xyz},
the CR species in each component $j$ should share a common cutoff
in rigidity $R_{j}$, which leads to $\gamma_{ij,\text{cut}}=(Z_{i}/m_{i})R_{j}$,
where $Z_{i}$ is the electric-charge of the CR species $i$. In Ref.~\citep{Gaisser:2013bla}
four different parametrization are found to be in good agreement with
the current UHECR data \cite{Nagano:1991jz,Glasmacher:1999id,Arqueros:1999uq,Takeda:2002at,Antoni:2005wq,Abbasi:2007sv,Amenomori:2008aa,Aartsen:2013wda,Ter-Antonyan:2014hea,Prosin:2016rqu,Fenu:2017hlc,Arteaga-Velazquez:2017rmh}.
We choose one of the ``Global-Fit'' parametrization with $n=3$.
The best-fit values of the rigidity cutoffs are $R_{1,2,3}=120$ TV,
4 PV and 1.3 EV, respectively \citep{Gaisser:2013bla}. Compared with
other parametrizations, this one is the most economic and conservative
as the final cutoff of $R_{3}$ is the lowest, which leads to the
lowest CRDM flux at high energy region. The details of the parametrizations
are summarized in Appendix-\ref{sec:Parametrizaitons-of-UHECR}. 

\subsection{DM flux from ultrahigh-energy CRs}

\begin{figure}
\includegraphics[width=0.75\columnwidth]{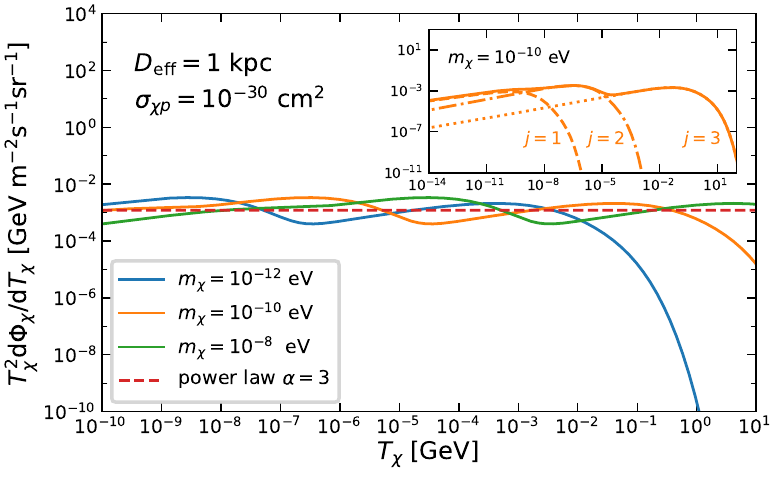}

\caption{\label{fig:CRDM-flux}(solid lines) CRDM flux rescaled by $T_{\chi}^{2}$
as a function of kinetic energy with cross section $\sigma_{\chi p}=10^{-30}\thinspace\text{cm}^{2}$
for different CRDM masses, . The spectral structure in the flux is
due to the three-component nature of the parametrization of CR \citep{Gaisser:2013bla}.
The inset shows the contribution from each individual CR component
($j=1,2,3$) for the case of $m_{\chi}=10^{-10}\thinspace\text{eV}$.
The simple case where the CR is proton dominant and follows a single
power-law with index $\alpha=3$ and $\Phi^{0}=2.6\times10^{2}\thinspace\text{cm}^{-2}\text{s}^{-1}\text{sr}^{-1}$
is also shown (horizontal dashed line) for comparison.}
 
\end{figure}

Fig.~\ref{fig:CRDM-flux} shows the CRDM flux calculated numerically
from Eqs.\,(\ref{eq:Tmax}) and (\ref{eq:CRDM-flux}) without using
approximations. In the calculation we take the dipole form factors
for light CR species H and He \citep{Perdrisat:2006hj}, and the Helm
form factor for heavier species \citep{Helm:1956zz,Lewin:1995rx}.
In the energy region where $T_{\chi}$ is far below the lowest cut
off, since $\alpha\approx2.7$ the CRDM fluxes follow an approximate
power law $\sim T_{\chi}^{-1.85}$ and scale with DM mass as $m_{\chi}^{-0.15}$
which is a rather weak $m_{\chi}$-dependence. Thus lighter CRDM particle
have slightly larger flux. In the cutoff dominated region, due to
the superposition of various cutoffs $\gamma_{ij,\text{cut}}$ in
the three components, the CRDM fluxes fluctuate around the power law
case of $T_{\chi}^{-2}$ over many orders of magnitude in kinetic
energy before reaching the last cutoff, and are insensitive to $m_{\chi}$,
which is expected from Eq.\,(\ref{eq:CRDM-flux-powerlaw}) and can
be clearly seen in Fig.~\ref{fig:CRDM-flux}. Above the final cutoff
$R_{3}$, the CRDM flux drops rapidly; the flux of lighter CRDM particle
drops faster, as expected from Eq.\,(\ref{eq:CRDM_cutoff}). Taking
the case of $m_{\chi}=10^{-10}\:\text{eV}$ as an example, the CR
protons in the three components lead to the induced CRDM flux with
three cut off at $T_{\chi,\text{cut}}^{\text{max}}=3.3\times10^{-9}$,
$3.6\times10^{-6}$ and $0.38\,\text{GeV}$, respectively, as can
be seen from the inset of Fig. \ref{fig:CRDM-flux}. If the cutoff
is too low, the CRDM cannot be energetic enough to produce enough
recoil energy to be detected by the DD experiments. Thus a lower limit
on $m_{\chi}$ for a given DD experiment exists.

\section{Earth attenuation\label{sec:Earth-atten} }

Before arriving at the underground detectors, CRDM may loss a non-negligible
fraction of energy due to the same elastic scatterings with nucleus
within the outer crust of Earth.  We adopt an analytic approach for
Earth attenuation based on average energy loss \citep{Kouvaris:2014lpa,Starkman:1990nj}.
 For simplicity, we only consider elastic scatterings which is an
irreducible process and neglect the effect of form factor. The decrease
of $T_{\chi}$ with depth $z$ due to the elastic scattering with
the nucleus $N$ in Earth's crust is given by
\begin{equation}
\frac{dT_{\chi}}{dz}=-\sum_{N}\left(\frac{\rho_{N}}{m_{N}}\right)\int_{0}^{T_{N}^{\text{max}}}T_{N}\frac{d\sigma_{\chi N}}{dT_{N}}dT_{N},\label{eq:dTdz}
\end{equation}
where $\rho_{N}$ is the mass density of nucleus $N$ in the crust,
and $T_{N}$ stands for the nucleus recoil energy which equals the
energy loss of the incident CRDM particle.\textcolor{black}{{} }Using
the expression of $T_{N}^{\text{max}}$ the energy loss can be approximated
as
\begin{equation}
\frac{dT_{\chi}}{dz}\approx-\kappa T_{\chi}^{2},
\end{equation}
where $\kappa=\sum_{N}\rho_{N}\sigma_{\chi N}/m_{N}^{2}$. We consider
the case where the scattering is isospin conserving, namely, $\sigma_{\chi n}\approx\sigma_{\chi p}$
such that the cross sections at nucleus and nucleon level are simply
related by $\sigma_{\chi N}\approx A_{N}^{2}\sigma_{\chi p}$ with
$A_{N}$ the nucleus mass number of $N$. We further adopt the relation
$m_{N}\approx A_{N}m_{p}$ which is a very good approximation. Under
these two simplifications, the factor $A_{N}$ cancels out in the
expression of $\kappa$, which gives $\kappa\approx\sigma_{\chi p}\rho_{\oplus}/m_{p}^{2}$
with $\rho_{\oplus}\approx2.7\thinspace\text{g\ensuremath{\cdot}cm}^{-3}$
the average mass density of Earth's outer crust \citep{Rudnick:2003xyz}.
After integrating Eq.\,(\ref{eq:dTdz}), the CRDM kinetic energy
$T_{\chi}^{z}$ at depth $z$ is related to that at surface as
\begin{equation}
\frac{T_{\chi}}{T_{\chi}^{z}}\approx\frac{1}{1-z\sigma_{\chi p}\rho_{\oplus}T_{\chi}^{z}/m_{p}^{2}}.\label{eq:Tz}
\end{equation}
Thus the effect of energy loss is independent of both $m_{\chi}$
and the chemical composition of the crust. For a small enough cross
section $\sigma_{\chi p}\ll m_{p}^{2}/(z\rho_{\oplus}T_{\chi}^{z})$,
one obtains $T_{\chi}^{z}\approx T_{\chi}$. In the opposite case
where $\sigma_{\chi p}$ is large enough, $T_{\chi}^{z}$ will stop
tracking $T_{\chi}$, and reach a maximal value $T_{\chi}^{z,\text{max}}\approx m_{p}^{2}/z\sigma_{\chi p}\rho_{\oplus}$
with increasing $T_{\chi}$. The appearance of $T_{\chi}^{z,\text{max}}$
is due to the rapid energy loss proportional to $T_{\chi}^{2}$ for
relativistic incident particles, which leads to a sharp cutoff in
the CRDM flux at depth $z$. The CRDM flux $d\Phi_{\chi}/dT_{\chi}^{z}$
at depth $z$ can be evaluated from that at surface $d\Phi_{\chi}/dT_{\chi}$
through the relation $d\Phi_{\chi}/dT_{\chi}^{z}=(d\Phi_{\chi}/dT_{\chi})(dT_{\chi}/dT_{\chi}^{z})$
. 

\begin{figure}
\includegraphics[width=0.75\columnwidth]{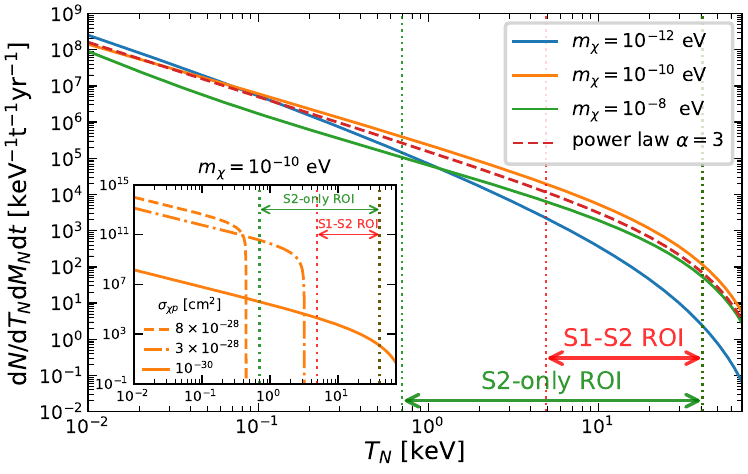}

\caption{\label{fig:recoil-spectrum} Recoil event rates of the scattering
between CRDM particles and target Xenon nuclei with different DM masses.
The value of $\sigma_{\chi p}$ and $D_{\text{eff}}$ are the same
as that in Fig. \ref{fig:CRDM-flux}. The inset shows the recoil rate
at different cross sections for $m_{\chi}=10^{-10}\thinspace\text{eV}$.
The vertical dotted lines indicate the region-of-interest (ROI) considered
by the Xenon-1T experiments for the S1-S2 \citep{Aprile:2018dbl}
and S2-only analyses \citep{Aprile:2019xxb}.}
\end{figure}

\section{Direct detection\label{sec:Direct-detection} }

\subsection{Recoil event spectrum}

The differential nuclear recoil event rate per target nucleus mass
$\Gamma=dN/dM_{N}dtdT_{N}$  of the $\chi N$ scattering at depth
$z$ is given by
\begin{equation}
\Gamma=\frac{4\pi\sigma_{\chi N}F^{2}(Q_{N}^{2})}{m_{N}}\int_{T_{\chi}^{z,\text{min}}(T_{N})}^{\infty}\frac{dT_{\chi}^{z}}{T_{N}^{\text{max}}(T_{\chi}^{z})}\frac{d\Phi_{\chi}}{dT_{\chi}^{z}},\label{eq:recoil}
\end{equation}
where $Q_{N}^{2}=2m_{N}T_{N}$. The scale of $T_{\chi}^{z}$ relevant
to direct detection is set by the lower limit of the integration $T_{\chi}^{z,\text{min}}(T_{N})\approx(T_{N}m_{N}/2)^{1/2}.$
For a detector located at depth $z\sim1\thinspace\text{km}$ and a
target nucleus mass $m_{N}\sim100\thinspace\text{GeV}$, in order
to produce a recoil energy $T_{N}$ which is close to the threshold
$T_{N}^{\text{thr}}\sim1\text{ keV}$, the required minimal $T_{\chi}^{z}$
is $\sim10\text{\thinspace MeV}.$ Thus the condition of $T_{\chi}^{z}\approx T_{\chi}$
leads to a typical requirement of $\sigma_{\chi p}\ll\mathcal{O}(10^{-27})\thinspace\text{cm}^{2}$.
Since the lower bounds of the excluded cross section is expected to
be much smaller $\sigma_{\chi p}\lesssim10^{-31}\thinspace\text{cm}^{2}$
\citep{Bringmann:2018cvk}, in deriving the lower bound of the excluded
$\sigma_{\chi p}$, the effect of Earth attenuation can be safely
neglected.  Using the CRDM flux from Eq. (\ref{eq:DM-flux-general}),
the recoil event rate can be written as
\begin{align*}
\text{\ensuremath{\Gamma}}= & \frac{\pi\sigma_{\chi N}\sigma_{\chi i}\rho_{\chi}^{\text{loc}}D_{\text{eff}}\Phi_{i}^{0}F^{2}}{2m_{\chi}^{3}\gamma_{i,\text{cut}}^{\alpha+3}}\left[-\Gamma(-3-\alpha_{i},t')+\frac{\Gamma(-1-\alpha_{i},t')}{t'^{2}}\right],
\end{align*}
where $t'=[m_{N}T_{N}/(8m_{\chi}^{2}\gamma_{i,\text{cut}}^{4})]^{1/4}$.
In the case where the CR flux follows a power law with index $\alpha_{i}$,
the recoil event rate can be obtained analytically from Eqs.\,(\ref{eq:CRDM-flux-powerlaw})
and (\ref{eq:recoil}) as follows
\begin{align}
\Gamma\approx & \frac{\pi\sigma_{\chi N}\sigma_{\chi i}\rho_{\chi}^{\text{loc}}D_{\text{eff}}\Phi_{i}^{0}F^{2}}{(1+\alpha_{i})(3+\alpha_{i})m_{N}^{3}}\left(\frac{m_{N}}{m_{\chi}}\right)^{\frac{3-\alpha_{i}}{2}}\left(\frac{T_{N}}{8m_{N}}\right)^{-\frac{3+\alpha_{i}}{4}},
\end{align}
which explicitly shows that if $\alpha_{i}\lesssim3$, the recoil
event rate is proportional to $T_{N}^{-3/2}$ and does not decrease
with a decreasing $m_{\chi}$. For the case $\alpha_{i}\approx3$,
the derived upper limit on $\sigma_{\chi p}$ will be insensitive
to $m_{\chi}$. In Fig. \ref{fig:recoil-spectrum}, we show the full
numerical results of recoil event rate of the scattering between the
CRDM particles and xenon nuclei. The approximate power-law behavior
of the recoil spectrum can be clearly seen for $T_{N}\lesssim\text{keV}$.
Above $\sim10\,\text{keV}$, the suppression due to the Helm form
factor is visible. Due to the power-law like spectrum of the recoil
event rate,  the experiment with lower threshold $T_{N}^{\text{thr}}$
will be more sensitive to lighter DM particles.

For large enough cross sections, the appearance of\textcolor{black}{{}
$T_{\chi}^{z,\text{max}}$ leads to a cutoff in the recoil event spectrum.
If the corresponding recoil energy is below the threshold $T_{N}^{\text{thr}}$,
it will form a blind spot for direct detection. This possibility is
illustrated in the inset of Fig. \ref{fig:recoil-spectrum}, which
will lead to $m_{\chi}$-independent upper bounds on the excluded
value of $\sigma_{\chi p}.$ }

\subsection{Xenon-1T detector response and data analysis}

We numerically calculate excluded regions in the $(m_{\chi},\sigma_{\chi p})$
plane at $90\%\text{ C.L.}$ for CRDM from the data of Xenon-1T experiment
located at depth $z\approx1.4\text{ km}$ \citep{Aprile:2018dbl,Aprile:2019xxb}.
The Xenon-1T experiment utilizes the liquid xenon time projection
chambers to detect the recoil energy of the target nuclei from their
scattering with DM particles. The deposited energy can produce a prompt
scintillation signal (S1) and ionization electrons which are extracted
into gaseous xenon and produce proportional scintillation light (S2).
The S2/S1 signal size ratio allows for discrimination between nuclear
recoils and electron recoils.  Since the nuclear recoil event rate
from the collisions with CRDM is quite different from that with the
nonrelativistic halo DM, to be more accurate on the effect of threshold,
we derive the limits directly from the distribution of the signals
of S1 and S2, rather than naively rescaling the reported experimental
limits from halo WIMP searches \citep{Bringmann:2018cvk,Dent:2019krz,Bondarenko:2019vrb,Cappiello:2019qsw}. 

 For the calculations from the deposited recoil energy $T_{N}$ to
the position-corrected signals cS1 and $\text{cS2}_{b}$, we closely
follow Ref.\,\citep{Aprile:2019dme}. For a deposited recoil energy
$T_{N}$, the averaged number of photons $\left\langle N_{\gamma}\right\rangle $
and charges $\left\langle N_{e}\right\rangle $ are given by
\begin{align}
\frac{\left\langle N_{\gamma}\right\rangle }{T_{N}} & =\frac{L}{W}\cdot\frac{\left\langle r\right\rangle +\left\langle N_{ex}/N_{i}\right\rangle }{1+\left\langle N_{ex}/N_{i}\right\rangle }\nonumber \\
\frac{\left\langle N_{e}\right\rangle }{T_{N}} & =\frac{L}{W}\cdot\frac{1-\left\langle r\right\rangle }{1+\left\langle N_{ex}/N_{i}\right\rangle },
\end{align}
where $W$ is the average energy required to create either an excitation
or ion-electron pair in the liquid xenon, $L$ is the Lindhard factor,
$N_{ex}/N_{i}$ is the excitation-to-ion ratio and $r$ is the recombination
probability.   The prompt and scintillation light hitting the PMT
photocathode will produce photoelectrons (PEs). The average number
of PEs observed per emitted photon is described by the gain factor
$g_{1}'(x,y,z)$, and the amplification factor for charge signal is
described by the parameter $g_{2}'(x,y)$. The S1 and S2 signal are
constructed from $N_{\text{pe}}$ and $N_{\text{prop}}$. The bias
and fluctuations are modeled by Gaussian distribution with Gauss($\delta_{\text{s1}},\Delta\delta_{\text{s1}}$)
and Gauss($\delta_{\text{s2}},\Delta\delta_{\text{s2}}$) respectively.
Finally, the spatial dependences of S1 and S2 will be corrected which
leads to the signal cS1 and $\text{cS2}_{b}$. The Xenon-1T collaboration
has performed the S1+S2 analysis based with an effective exposure
of 1 ton-yr. In the analysis the DM search was performed between $3<\text{cS1}<70\,\text{PE}$,
corresponding to an average $\text{keV}_{\text{nr}}$ of $4.9-40.9\,\text{keV}$
with an effective exposure of one ton-year\citep{Aprile:2018dbl}
. The Xenon-1T collaboration also formed the S2-only analysis with
$\text{cS2}_{b}>150\,\text{PE}$, corresponding to a threshold of
$0.7\thinspace\text{keV}_{\text{nr}}$ with an effective exposure
of 22 ton-days \citep{Aprile:2019xxb}.

We calculate the signal distributions of the scattering between CRDM
particles and target Xenon nuclei, and derive the excluded regions
in $(m_{\chi},\sigma_{\chi p})$ plane for the Xenon-1T data (S1+S2)
using the binned Poisson statistic approach \citep{Green:2001xy,Savage:2008er}.
The distribution of the background events are taken from the Xenon-1T
analysis. The calculation procedure, main parameters and extended
results with different parametrizations of CR flux are summarized
in appendix-\ref{sec:Xenon-1T-data-analysis}.
\begin{figure}
\includegraphics[width=0.75\columnwidth]{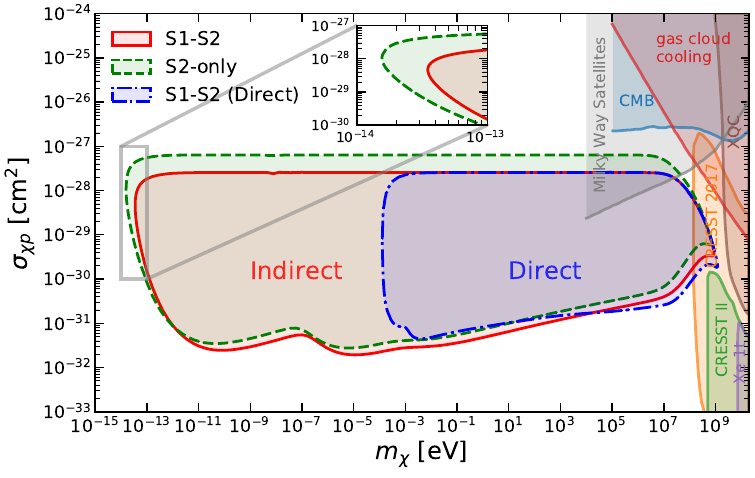}

\caption{\label{fig:Limits} Exclusion regions in the $(m_{\chi},\sigma_{\chi p})$
plane for CRDM. The result using UHECR indirectly measured by ground-based
experiments \citep{Gaisser:2013bla} and the Xenon-1T S1-S2 \citep{Aprile:2018dbl}
(S2-only \citep{Aprile:2019xxb}) data is shown as a solid (dashed)
contour. For comparison, the results using CR directly measured by
space-based experiments \citep{DellaTorre:2016jjf} with a maximal
energy of 200~TeV is also shown as a dot-dashed contour. A selection
of constraints on halo DM from other experiments such as Xenon-1T
\citep{Aprile:2018dbl}, CRESST-II \citep{Angloher:2015ewa}, CRESST
surface run \citep{Angloher:2017sxg}, XQC \citep{Mahdawi:2018euy}
and CMB \citep{Xu:2018efh}, gas cloud cooling \citep{Bhoonah:2018wmw},
Milky way satellite population \citep{Nadler:2019zrb} are also shown.}
\end{figure}

\subsection{Results}

In Fig.\,\ref{fig:Limits}, we show the final constraints from analysing
the Xenon-1T data. It can be seen from the figure that the lower bound
of the excluded region reaches $\sigma_{\chi p}\lesssim10^{-(32-31)}\ \text{cm}^{2}$
can be extended down to DM particle mass $\sim10^{-12}\text{ eV}$.
In the sub-eV region, the shape of the excluded region is directly
related to the structures in the UHECR flux. The most stringent limits
of $\sigma_{\chi p}\lesssim3\times10^{-32}\thinspace\text{cm}^{2}$
at $m_{\chi}\sim10^{-11}\thinspace\text{eV}$ and $10^{-5}\thinspace\text{eV}$
correspond to the ``knee'' and the ``toe'' structures of the UHECR
flux. The exclusion region closes at $m_{\chi}\sim10^{-14}\ \text{eV}$,
which corresponds to the observed suppression of the UHECR flux at
$\sim10^{20}\thinspace\text{eV}$ possibly due to the GZK cutoff.
For comparison purpose, in Fig. \ref{fig:Limits} we also show the
results using the CR proton and Helium fluxes in \citep{DellaTorre:2016jjf}
which are based on the space-based direct CR measurements and are
only available for energy below 200~TeV \citep{Aguilar:2015ooa,Aguilar:2017hno,Yoon:2011aa}.
It can be seen that using the ground-based indirect UHECR measurements
can extend the previous constraints by about ten orders of magnitude
towards lower DM particle mass. The constraints are conservative,
as we adopted the ``Global-Fit'' parametrization of the UHECR flux.
For other parametrizations such as ``H3a'', ``H4a'' and ``Global-Fit4'',
the resulting constraints are even stronger towards lower DM mass,
as it is shown in Fig. \ref{fig:Exclusion-regions-more} of Appendex-\ref{sec:Excluded-regions-Appdx}. 

As it can be seen in the inset of Fig.\,\ref{fig:Limits}, we also
find that the constraints on the lowest CRDM mass is very sensitive
to the detector threshold as the recoil event spectrum from CRDM is
quite different from halo DM. Due to the lower threshold $\sim0.7\thinspace\text{keV}_{\text{nr}}$
of the S2-only data \citep{Aprile:2019xxb}, the constraints from
the S2-only data which has an exposure about an order of magnitude
smaller than that of the S1-S2 data (with a threshold of $\sim4.9\thinspace\text{keV}_{\text{nr}}$
\citep{Aprile:2018dbl}) turns out to be more sensitive to lighter
CRDM below $10^{-12}\thinspace\text{eV}$.

In the simple analytic approach adopted in this work, the upper bound
of the excluded region due to the Earth attenuation is found to be
$\sigma_{\chi p}\lesssim8\times10^{-28}\thinspace\text{cm}^{2}$,
and is almost insensitive to $m_{\chi}$, as expected from the fast
energy loss proportional to $T_{\chi}^{2}$ in Earth attenuation discussed
in Sec. \ref{sec:Earth-atten}. 

\section{Discussions/Conclusions\label{sec:Conclusions} }

 In summary, we have derived novel constraints on ultralight DM boosted
by UHECRs. The constraints obtained in this work are highly model
independent, as only the DM-nucleon scattering cross setion is relevant.
The approach only requires the information of the present-day local
Universe, even the standard cosmology is not assumed. Thus the constraints
are complementary to that derived from different epochs of the Universe,
such as the observations of CMB \citep{Slatyer:2018aqg,Gluscevic:2017ywp,Boddy:2018kfv,Xu:2018efh},
Lyman-$\alpha$ forest \citep{Murgia:2018now} and 21 cm radiations
\citep{Slatyer:2018aqg}, etc.. If the ultralight DM particles reached
chemical and kinetic equilibrium in the early Universe, very stringent
constraints will arise from BBN and CMB. However, the conditions for
reaching thermal equilibrium require more information on the cross
section of DM annihilation and production process, which are in general
model dependent. For ultralight DM particles, annihilating into most
standard model particles are kinematically forbidden (for the analysis
try to connect DM scattering and annihilation cross sections using
crossing symmetry, see, e.g. \citep{Krnjaic:2019dzc}). 

In this work, we considered the simplest case where the scattering
cross section is a constant, i.e., energy and momentum-transfer independent.
For relativistic scatterings, it more likely that the differential
scattering cross section depends on the energy of both the incident
and outgoing particles. For some simple models, such as the fermionic
DM with a scalar mediator, it has been shown that the differential
scattering cross section can be greatly enhanced at high momentum
transfer. Consequently, the constraints on the total cross section
at the zero momentum transfer can be many orders of magnitude stronger
for lighter DM particles \citep{Bondarenko:2019vrb}. The approach
proposed in this work can be extended to the case with energy-dependent
cross sections in a straight forward manner.

\begin{acknowledgments}
We are grateful to Qian Yue and Li-Tao Yang for helpful discussions.
This work is supports in part by The National Key R\&D Program of
China No.~2017YFA0402204, the National Natural Science Foundation
of China (NSFC) No.~11825506, No.~11821505, No.~11851303, No.~11947302,
and the Key Research Program of the Chinese Academy of Sciences, No.
XDPB15.
\end{acknowledgments}

\appendix

\section{Parametrizaitons of UHECR flux\label{sec:Parametrizaitons-of-UHECR}}

CR particles with energy above a few hundred TeV are mainly measured
by the ground-based air-shower arrays which detect the cascades of
secondary particles from the interactions of primary CR particles
in the Earth atmosphere. In such indirect experiments, the information
about the chemical composition is limited to determining the relative
abundance of the main species or groups. Thus, it is likely that there
exists different parametrizations which can describe the data equally
well. We take the $n$-component power-law parametrization in which
the CR total energy spectrum of the CR species $i$ has the following
form~\citep{Gaisser:2013bla} 
\begin{equation}
\frac{\mathrm{d}\Phi_{i}^{\mathrm{LIS}}}{\mathrm{d}E_{i}}=\sum_{j=1}^{n}c_{ij}E_{i}^{-\alpha_{ij}}\exp\left[-\frac{E_{i}}{Z_{i}R_{j}}\right],\label{eq:Hillas-1}
\end{equation}
where $j=(1,\dots,n$) is the component index, $E_{i}$ (in unit of
GeV) is the total energy of CR species $i$. The normalization constants
$c_{ij}$ are related to $\Phi_{ij}^{0}$ in Eq. (5) of the main text
by $\Phi_{ij}^{0}=m_{i}^{1-\alpha_{ij}}c_{ij}$ where $m_{i}$ (in
unit of GeV) is the mass of CR species $i$. A global analysis to
the recent UHECR data has been performed in~\citep{Gaisser:2013bla}.
We adopt one of the three-component ``Global-Fit'' model as the
benchmark model with the parameters listed in Table~\ref{tab:globalfit}.
In this parametrization, the first rigidity cutoff $R_{i}$ is around
100 TV which is the typical maximal energy from the acceleration of
SNR with magnetic field around a few $\mu$ Gauss. It also well reproduce
the observed hardening in the CR all-particle spectrum above 200 GeV~\citep{Ahn:2010gv,Adriani:2011cu}.
In the figure, we also list a slightly extended four-component parametrization
(referred to as ``Global-Fit4'').

\begin{table}[htb]
\centering %
\begin{tabular}{llccccccc}
\hline 
 &  & p & He & C & O & Fe & 50\textless Z\textless 56 & 78\textless Z\textless 82\tabularnewline
\hline 
$R_{1}=120$ TV & $c_{i1}$ & 7000 & 3200 & 100 & 130 & 60 &  & \tabularnewline
 & $\alpha_{i1}$ & 2.66 & 2.58 & 2.4 & 2.4 & 2.3 &  & \tabularnewline
\hline 
$R_{2}=4$ PV & $c_{i2}$ & 150 & 65 & 6 & 7 & 2.3 (2.1) & 0.1 & 0.4 (0.53)\tabularnewline
 & $\alpha_{i2}$ & 2.4 & 2.3 & 2.3 & 2.3 & 2.2 & 2.2 & 2.2\tabularnewline
\hline 
$R_{3}=1.3\ (1.5)$ EV & $c_{i3}$ & 14 (12) &  &  &  & 0.025 (0.011) &  & \tabularnewline
 & $\alpha_{i3}$ & 2.4 &  &  &  & 2.2 &  & \tabularnewline
\hline 
$(R_{4}=40\text{ EV})$ & $c_{i4}$ & (1.2) &  &  &  &  &  & \tabularnewline
 & $\alpha_{i4}$ & (2.4) &  &  &  &  &  & \tabularnewline
\hline 
\end{tabular}\caption{Normalization constants $c_{ij}$, power indexes $\alpha_{ij}$, and
rigidity cutoffs $R_{j}$ in the parametrization of ``Global-Fit''
and ``Global-Fit4'' in~\citep{Gaisser:2013bla}. The parameters
of ``Global-Fit4'' which are different from those of the ``Global-Fit''
are shown in parentheses.}
\label{tab:globalfit}
\end{table}

Two alternative parametrizations~\citep{Gaisser:2012zz} based on
the Hillas model~\citep{Hillas:2005cs} are also found in good agreement
with data, which are labled as ``H3a'' and ``H4a'' in~\citep{Gaisser:2013bla}.
The major difference from the ``Global-Fit'' parametrization is
that the first rigidity cutoff is quite high about 4 PV, which is
responsible for the ``knee'' structure. In this type of parametrization
the ``ankle'' represent the transition between the galactic and
extra-galactic contributions. The corresponding parameters are listed
in Table~\ref{tab:h3ah4a}.

\begin{table}[tb]
\centering

\begin{tabular}{llccccccc}
\hline 
 &  & p & He & CNO & Mg-Si & Fe &  & \tabularnewline
\hline 
$R_{1}=4$ PV & $c_{i1}$ & 7860 & 3550 & 2200 & 1430 & 2120 &  & \tabularnewline
 & $\alpha_{i1}$ & 2.66 & 2.58 & 2.63 & 2.67 & 2.63 &  & \tabularnewline
\hline 
$R_{2}=30$ PV & $c_{i2}$ & 20 & 20 & 13.4 & 13.4 & 13.4 &  & \tabularnewline
 & $\alpha_{i2}$ & 2.4 & 2.4 & 2.4 & 2.4 & 2.4 &  & \tabularnewline
\hline 
$R_{3}=2\ (60)$ EV & $c_{i3}$ & 1.7 (200) & 1.7 (0.0) & 1.14 (0.0) & 1.14 (0.0) & 1.14 (0.0) &  & \tabularnewline
 & $\alpha_{i3}$ & 2.4 (2.6) & 2.4 & 2.4 & 2.4 & 2.4 &  & \tabularnewline
\hline 
\end{tabular}\caption{Same as Table~\ref{tab:globalfit}, but for parametrizations of ``H3a''
and ``H4a'' in~\citep{Gaisser:2013bla}. The parameters of ``H4a''
which are different from those of ``H3a'' are shown in parentheses.}
\label{tab:h3ah4a}
\end{table}

\begin{figure}[bth]
\centering \includegraphics[width=0.45\textwidth]{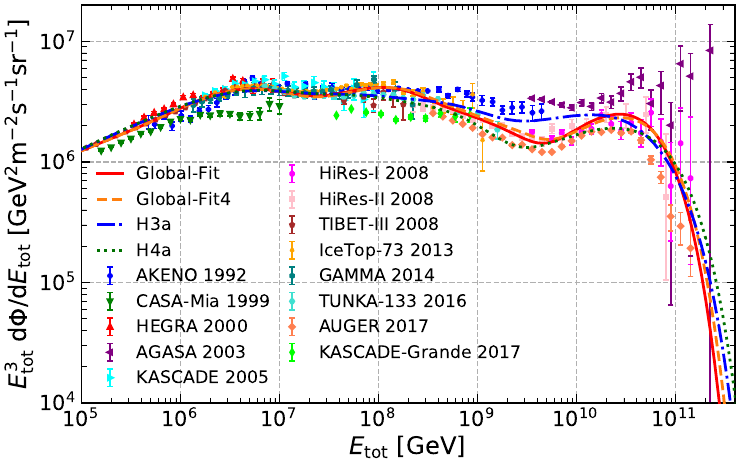}
\includegraphics[width=0.45\textwidth]{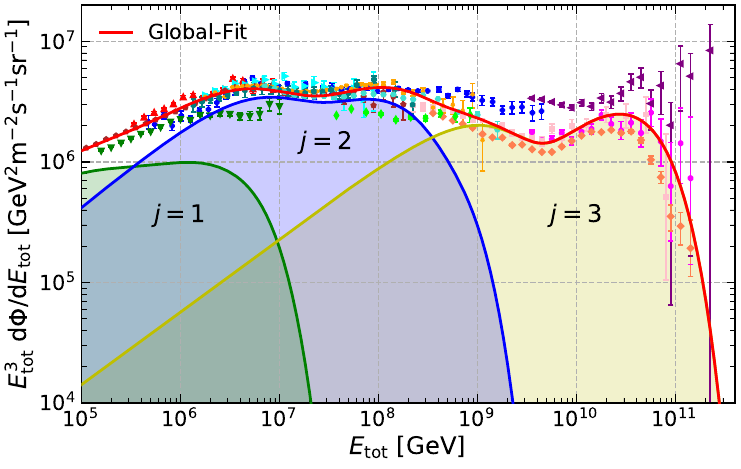} \caption{(Left) CR all-particle spectra for four type of parametrizations in
\citep{Gaisser:2013bla} together with the current experimental data
\cite{Nagano:1991jz,Glasmacher:1999id,Arqueros:1999uq,Takeda:2002at,Antoni:2005wq,Abbasi:2007sv,Amenomori:2008aa,Aartsen:2013wda,Ter-Antonyan:2014hea,Prosin:2016rqu,Fenu:2017hlc,Arteaga-Velazquez:2017rmh}.
Right) Contributions from the three individual components in the
``Global-Fit" parametrization in \citep{Gaisser:2013bla}}
\label{fig:flux_Hillas}
\end{figure}

The CR all-particle fluxes of the four parametrizations are shown
in Figure~\ref{fig:flux_Hillas} together with the recent experiments
data.

\section{Xenon-1T data analysis\label{sec:Xenon-1T-data-analysis}}


For the data analysis of the Xenon-1T experiment, we adopt the signal
response model described by the Xenon-1T collaboration in \citep{Aprile:2019dme}.
The Xenon-1T experiment utilizes the liquid xenon time projection
chambers to detect the recoil energy of the target nuclei from the
scattering with DM particles. The deposited energy can produce a prompt
scintillation light signal (S1), and the ionization electrons extracted
from liquid xenon into gaseous xenon can produce proportional scintillation
light (S2). The S2/S1 signal size ratio allows for discrimination
between nuclear recoil (NR) and electron recoil (ER) events. For a
deposited recoil energy $T_{N}$, the produced total number of quantum
$N_{q}$ is the sum of the number of excitons $N_{\text{ex}}$ and
ion-electron pairs $N_{i}$, which follows a binomial distribution
$N_{q}\sim\text{Binom}(T_{N}/W,L)$ where $W=13.8\ {\rm eV}$ is the
average energy required to create either an exciton or ion-electron
pair in the liquid xenon, and $L$ is the Lindhard factor. In the
case of NR, it is given by \citep{Lenardo:2014cva} 
\begin{equation}
L=\frac{k\ g(\epsilon)}{1+k\ g(\epsilon)},
\end{equation}
where $k$ is a normalization factor, the function $g(\epsilon)$
is proportional to the ratio of electric and nuclear stopping power,
which can be parametrized as $g(\epsilon)=3\epsilon^{0.15}+0.7\epsilon^{0.6}+\epsilon$,
where $\epsilon=11.5(T_{N}/{\rm keV})Z^{-7/3}$ with $Z=54$ the atomic
number of the xenon nucleus. The distribution of $N_{i}$ is described
by a binomial distribution $N_{i}\sim\text{Binom}(N_{q},1/(1+\left\langle N_{\text{ex}}/N_{i}\right\rangle ))$,
where $\langle N_{ex}/N_{i}\rangle$ is the averaged exciton-to-ion
ratio. The number of excitons is given by $N_{\text{ex}}=N_{q}-N_{i}$.
The excitons contribute to scintillation photon signals through de-excitation
process. The ionized electrons have a probability of $r$ to be recombined
into xenon atoms and produce scintillation photons, and a probability
of $(1-r)$ to escape the ion-electron pair. Thus the number of escaped
electrons is given by $N_{e}\sim\text{Binom}(N_{i},1-r)$ and the
total number of photons is $N_{\gamma}=N_{\text{ex}}+N_{i}-N_{e}$.
The recombination probability $r$ is modeled by a Gaussian distribution
$r\sim\text{Gauss(\ensuremath{\langle r\rangle,\Delta r})}$, where
$\langle r\rangle$ is the mean value and $\Delta r$ is the variance.
The mean value is calculated using the Thomas-Imel box model~\citep{Thomas:1987zz,Szydagis:2011tk}
\begin{equation}
\begin{aligned}\langle r\rangle & =1-\frac{\ln(1+N_{i}\varsigma/4)}{N_{i}\varsigma/4},\end{aligned}
\label{e3-1}
\end{equation}
where $\varsigma=0.057F^{-0.12}$ with $F=81~\text{V/cm}$ the electric
field. The variance is described by $\Delta r=q_{2}(1-e^{-T_{N}/q_{3}})$
with $q_{2}=0.034$ and $q_{3}=1.7$. In summary, the averaged number
of photons $\left\langle N_{\gamma}\right\rangle $ and charges $\left\langle N_{e}\right\rangle $
are given by 
\begin{align}
\frac{\left\langle N_{\gamma}\right\rangle }{T_{N}} & =\frac{L}{W}\cdot\frac{\left\langle r\right\rangle +\left\langle N_{\text{ex}}/N_{i}\right\rangle }{1+\left\langle N_{\text{ex}}/N_{i}\right\rangle },\nonumber \\
\frac{\left\langle N_{e}\right\rangle }{T_{N}} & =\frac{L}{W}\cdot\frac{1-\left\langle r\right\rangle }{1+\left\langle N_{\text{ex}}/N_{i}\right\rangle }.
\end{align}

The photon and charge signals are converted into photoelectron (PE)
emission of the photomultiplier tube (PMT) photocathode. The corresponding
gain factors are: $g_{1}'(x,y,z)$ the average number of PEs observed
per emitted photon, and $g_{2}'(x,y)$ the amplification factor for
charge signals. Both are spatial dependent.

\begin{figure}
\centering \includegraphics[width=0.45\textwidth]{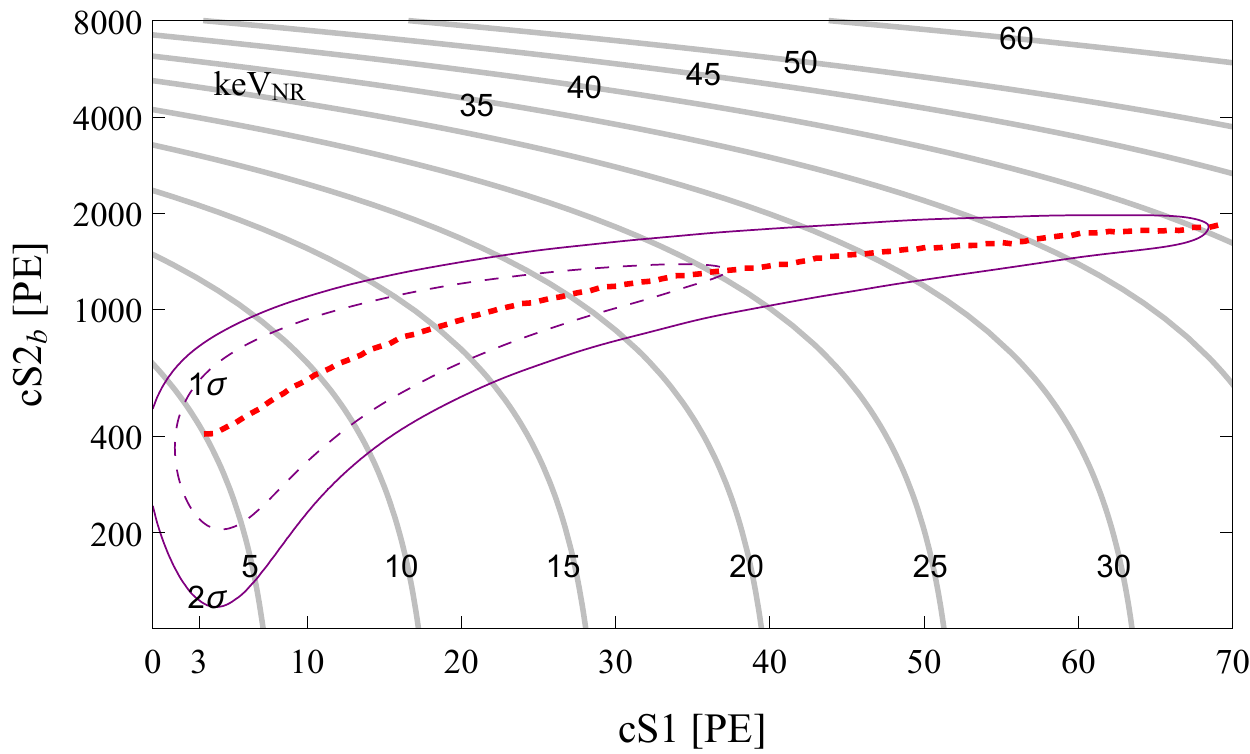} \includegraphics[width=0.45\textwidth]{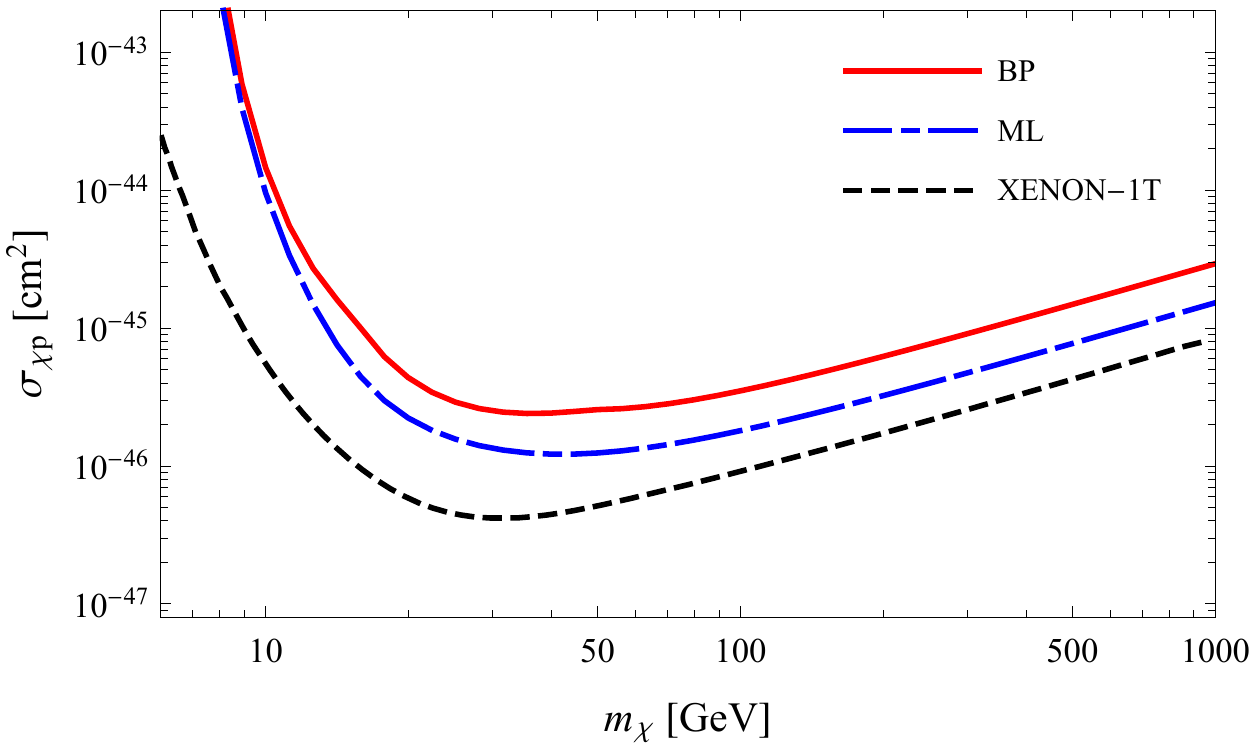}
\caption{Left) Nuclear recoil signal distributions from halo DM-Xe scattering
with $m_{\chi}=200\ {\rm GeV}$ and $\sigma_{\chi p}=4.7\times10^{-47}\ {\rm cm^{2}}$,
through MC simulation. The purple dashed and purple solid are $1\sigma$
and $2\sigma$ percentiles of DM signal, and the central value (red
dashed) is shown for reference. The gray lines are the iso-energy
contours in $\text{keV}_{\text{n}r}$. Right) 90\% C.L. upper limit
on $\sigma_{\chi p}$ from DM of SHM by two different statistic approaches
described above, binned Poisson (BP, red solid) maximum likelihood
(ML, blue dot-dashed). We also show the result from XENON-1T (black
dashed) for comparison.}
\label{fig-dis}
\end{figure}

The spatial-dependence of the signals S1 and S2 are corrected, which
results in the corrected signals $c\text{S1}$ and $c\text{S2}_{b}$
(corresponding to the S2 signals from the bottom PMTs). These two
quantities can be understood as spatial-averaged signals. For simplicity
we use the spatial-averaged gain factors of $g_{1}=0.142$ and $g_{2}=11.4$
for $c\text{S1}$ and $c\text{S2}_{b}$, respectively. Thus in this
case the number of PE is given by $N_{\text{pe}}\sim\text{Binom}(N_{\gamma},g_{1})$
and that of proportional signal is given by $N_{\text{prop}}\sim\text{Gauss}(N_{e}g_{2},\sqrt{N_{e}}\Delta g_{2})$,
with $\Delta g_{2}/g_{2}=0.25$. The $c\text{S1}$ and $c\text{S2}_{b}$
signals are constructed from $N_{\text{pe}}$ and $N_{\text{prop}}$.
The biases and fluctuations in the construction process is modeled
as 
\begin{align}
c\text{S1}/N_{\text{pe}}-1 & \sim\text{Gauss(\ensuremath{\delta_{s1},\Delta\delta_{s1}})},\\
c\text{S2}_{b}/N_{\text{prop}}-1 & \sim\text{Gauss}(\ensuremath{\delta_{s2},\Delta\delta_{s2}}),
\end{align}
where we adopt mean values of $\delta_{s1(s2)}=-0.065\,(0.034),$
and variances $\Delta\delta_{s1(s2)}=0.110\,(0.030)$. 
\begin{equation}
\begin{aligned}\langle cS1\rangle & \approx\frac{T_{N}\cdot L}{W}\frac{\langle r\rangle+\langle N_{ex}/N_{i}\rangle}{1+\langle N_{ex}/N_{i}\rangle}g_{1}\cdot(1+\delta_{s1}),\\
\langle cS2_{b}\rangle & \approx\frac{T_{N}\cdot L}{W}\frac{1-\langle r\rangle}{1+\langle N_{ex}/N_{i}\rangle}g_{2}\cdot(1+\delta_{s2}).
\end{aligned}
\label{e1-1}
\end{equation}
In the left panel of Fig.~\ref{fig-dis}, we show the Monte Carlo
(MC) simulation of the signal distributions from a typical scattering
between non-relativistic halo DM and xenon nucleus for $m_{\chi}=200\ {\rm GeV}$
and $\sigma_{\chi p}=4.7\times10^{-47}\ {\rm cm^{2}}$. We adopt the
Maxwellian distribution of the standard halo model (SHM) for DM with
the most probable velocity $v_{0}=220\ {\rm km/s}$, the escape velocity
$v_{esc}=544\ {\rm km/s}$, the local DM density $\rho_{\chi}^{{\rm loc}}=0.3\ {\rm GeV/cm^{3}}$
and the Earth velocity $v_{E}=232\ {\rm km/s}$~\citep{Drukier:1986tm}.
We assume the scattering process is elastic, spin-independent and
isospin-conserving, and adopt the Helm form factor~\citep{Lewin:1995rx}.
We find that the figure is in reasonable agreement with the result
of the Xenon-1T collaboration \citep{Aprile:2018dbl}.


For deriving the constraints on $\sigma_{\chi p}$ from halo DM, we
consider two different two statistic methods. The first one is the
Binned Poisson (BP) method \citep{Green:2001xy,Savage:2008er}. First,
Let us consider the single-bin case. Given an expectation value of
$\lambda=b+s$ events with $s$ the theoretical prediction and $b$
the expected background, the probability of observing $N_{{\rm obs}}$
events is given by the Poisson distribution 
\begin{equation}
P(N_{obs}|\lambda)=\text{Poiss}(N_{obs}|\lambda).
\end{equation}
The value of $\lambda_{p}$ at which $P(\lambda>\lambda_{p}|N_{obs})$
is smaller than $\alpha$ is excluded at $1-\alpha$ confidence level
(C.L.). For example, the 90$\%$~C.L. exclusion limit means $\alpha=0.1$.
The required $\lambda_{p}$ can be obtained from $P(\lambda>\lambda_{p}|N_{obs})=P(N<N_{obs}|\lambda_{p})<\alpha$.
In the case of multiple bins, if $(1-\alpha_{{\rm bin}})$ is the
probability of seeing $\lambda<\lambda_{p}$ in that bin, the possibility
$(1-\alpha)$ of seeing $\lambda<\lambda_{p}$ in any of the bins
is given by the binomial distribution 
\begin{equation}
1-\alpha=(1-\alpha_{{\rm bin}})^{N_{{\rm bin}}},
\end{equation}
where $N_{{\rm bin}}$ is the number of bins. For a desired exclusion
level of $1-\alpha$, we then use this relation to determine $\alpha_{{\rm bin}}$,
and find the value of $\lambda_{p}$. 

The second one is the Maximal Likelihood (ML) method. In this method,
the joint likelihood is obtained by the product of individual likelihoods
in each bin $i$, i.e., $\mathcal{L}=\prod_{i}\mathcal{L}_{i}$ where
$\mathcal{L}_{i}=\text{Poiss}(N_{obs}|\lambda_{i})$ is the Poisson
distribution. The theoretical prediction of the event number depends
on DM parameters, e.g. $\lambda_{i}=\lambda_{i}(m_{\chi},\sigma_{\chi p})$
The test statistics is defined as 
\begin{align}
\text{TS}=-2\ln\frac{\mathcal{L}(m_{\chi},\sigma_{\chi p})}{\mathcal{L}(\hat{m}_{\chi},\hat{\sigma}_{\chi p})},
\end{align}
where $\hat{m}_{\chi}$ and $\hat{\sigma}_{\chi p}$ are the best-fit
DM parameters which maximize the likelihood. For a given value of
$m_{\chi}$, the TS should approximately follow a $\chi^{2}$-distribution
with one degree-of-freedom \citep{Rolke:2004mj}. 
The value of $\sigma_{\chi}$ for which $\text{TS}>2.7$ are excluded
at 90$\%$ C.L..

For the S1-S2 combined data analysis, the Xenon-1T collaboration adopted
the energy regions of interest (ROI) for cS1 as $3<\text{cS1}<70$~PE,
corresponding to an average energy of 
4.9\textendash 40.9 $\text{keV}_{\text{nr}}$. The ROI for $\text{cS2}_{b}$
is $50.1<\text{cS2}_{b}<7940$~PE. The selection of ROIs affects
the total acceptence. We take the total acceptance due to the data
selection, reconstruction, noise rejection, S1-S2 correlation and
single scattering, etc.. from Fig.~14 of \citep{Aprile:2019bbb}.
In deriving the constraints on CRDM, we use the data of $\text{cS2}_{b}$
distribution shown in Fig.~4 of \citep{Aprile:2018dbl}. In the figure
the distribution is shown with respect to the rescaled quantity $(\text{cS2}_{b}-\mu_{\text{ER}})/\sigma_{\text{ER}},$
where $\mu_{\text{ER}}$ and $\sigma_{\text{ER}}$ are the ER mean
and $1\sigma$ quantile, respectively. We take $\mu_{\text{ER}}=1958\ {\rm PE}$
and $\sigma_{\text{ER}}=408\ {\rm PE}$ from the $\text{cS2}_{b}$
distribution shown in Fig.~3 of~\citep{Aprile:2018dbl}. The number
of signal counts $s_{i}$ in a given bin of $\text{cS2}_{b}$ signal
is given by 
\begin{equation}
s_{i}=E_{{\rm X}}\int_{(cS2_{b})_{i}^{\text{low}}}^{(cS2_{b})_{i}^{\text{up}}}d\langle cS2_{b}\rangle\frac{dN}{dT_{N}}\frac{dT_{N}}{d\langle cS2_{b}\rangle}\epsilon_{2}(\langle cS2_{b}\rangle),
\end{equation}
where $E_{{\rm X}}=1{\rm \ tone\cdot year}$ is the total exposure
of the XENON-1T data, $(cS2_{b})_{i}^{\text{low(up)}}$ is the lower
(upper) endpoints of the $i$-th bin of the corresponding $cS2_{b}$
signal. $\epsilon_{2}=A_{1}(cS1)A_{2}(cS2_{b})$ is the total efficiency
of $cS2_{b}$, 
where $A_{1,2}$ are the acceptance within the ROI of $cS1$ and $cS2_{b}$,
respectively, and are vanishing outside the ROIs. The value of $\langle cS1\rangle$
and $\langle cS2_{b}\rangle$ are related to each other through Eq.~\ref{e1-1},
so the total efficiency of S2 can be written as a function of $\langle cS2_{b}\rangle$
only. For the background event number $b_{i}$, we directly adopt
the overall expected background given by XENON-1T, which include electric
recoils, neutron, surface, accidental coincidence (AC), and coherent
elastic neutrino-nucleus scatters (${\rm CE\nu NS}$).

In the right panel of Fig.~\ref{fig-dis} we show the upper limits
on $\sigma_{\chi p}$ at 90\% C.L. for SHM DM using the BP and ML
statistic methods. The DM local density and velocity distribution
are the same as that for Fig.~\ref{fig-dis}. In the figure, the
limits obtained by the Xenon-1T collaboration using a full profile
likelihood analysis is also shown for comparison. It can be seen that
the limits obtained in BP and ML approaches are quite conservative.
As we are interested in conservative constraints on CRDM properties,
we adopt the BP statistic approach in the main text.


Using the relation between the S2 signal and the averaged recoil energy
for NR and ER process shown in Fig.~6 of the supplementary material
of~\citep{Aprile:2019xxb}, we convert the Xenon-1T S2-only data
in Fig.~4 Ref.~\citep{Aprile:2019xxb} into the NR recoil energy
distribution $dN/dT_{N}$. 
The total number of events in the $i$-th energy bin is given by For
$s_{i}$, we have 
\begin{equation}
s_{i}=\int_{(T_{N})_{i}^{\text{low}}}^{(T_{N})_{i}^{\text{up}}}dT_{N}\frac{dN}{dt_{N}}\epsilon_{{\rm ex}}(T_{N}),
\end{equation}
where $(T_{N})_{i}^{\text{low(up)}}$ is the lower (upper) endpoints
of the $i$-th energy bin, and $\epsilon_{{\rm ex}}(T_{N})$ is the
effective exposure which is a function of recoil energy $T_{N}$ shown
in Fig.~1 of Ref.~\citep{Aprile:2019xxb}. For the background event
number, we use the expected background given by XENON-1T, which include
cathode, CEvNS, flat ER background for S2-only data.

\section{Excluded regions for CRDM \label{sec:Excluded-regions-Appdx} }

\begin{figure}[tbh]
\centering \includegraphics[width=0.8\linewidth]{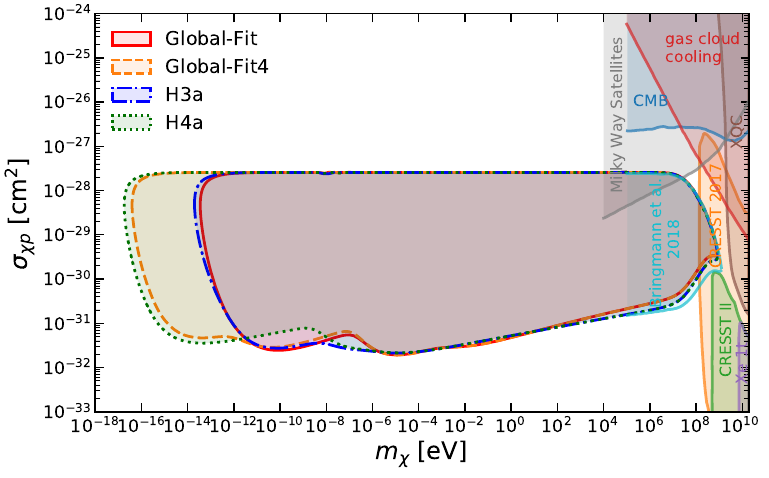}\caption{Exclusion regions in the ($m_{\chi}$, $\sigma_{\chi p}$) plane from
the Xenon-1T data on the S1-S2 signals and four CR parameterizations\label{fig:Exclusion-regions-more}.}
\label{fig:limits_compare}
\end{figure}

Making use of the Xenon-1T S1-S2 data, we derive the excluded regions
in $(m_{\chi},\sigma_{\chi p})$ plane using the BP statistic approach
for the four different parametrizations of the primary CR flux. The
results are shown in Figure~\ref{fig:limits_compare}. It can be
seen that the constraints based on the ``Global-Fit" parametrization
is quite conservative compared with other parametrizations.

\bibliographystyle{arxivref}
\bibliography{crdm_ref_spire,misc,supplementary_material}

\providecommand{\href}[2]{#2}\begingroup\raggedright\begin{thebibliography}{10}

\bibitem{Kouvaris:2016afs}
C.~Kouvaris and J.~Pradler, ``{Probing sub-GeV Dark Matter with conventional
  detectors},'' \href{http://dx.doi.org/10.1103/PhysRevLett.118.031803}{{\em
  Phys. Rev. Lett.} {\bfseries  118} no.~3, (2017) 031803},
  \href{http://arxiv.org/abs/1607.01789}{{\ttfamily arXiv:1607.01789
  [hep-ph]}}.

\bibitem{Ibe:2017yqa}
M.~Ibe, W.~Nakano, Y.~Shoji, and K.~Suzuki, ``{Migdal Effect in Dark Matter
  Direct Detection Experiments},''
  \href{http://dx.doi.org/10.1007/JHEP03(2018)194}{{\em JHEP} {\bfseries  03}
  (2018) 194}, \href{http://arxiv.org/abs/1707.07258}{{\ttfamily
  arXiv:1707.07258 [hep-ph]}}.

\bibitem{Dolan:2017xbu}
M.~J. Dolan, F.~Kahlhoefer, and C.~McCabe, ``{Directly detecting sub-GeV dark
  matter with electrons from nuclear scattering},''
  \href{http://dx.doi.org/10.1103/PhysRevLett.121.101801}{{\em Phys. Rev.
  Lett.} {\bfseries  121} no.~10, (2018) 101801},
  \href{http://arxiv.org/abs/1711.09906}{{\ttfamily arXiv:1711.09906
  [hep-ph]}}.

\bibitem{Gluscevic:2017ywp}
V.~Gluscevic and K.~K. Boddy, ``{Constraints on Scattering of keV--TeV Dark
  Matter with Protons in the Early Universe},''
  \href{http://dx.doi.org/10.1103/PhysRevLett.121.081301}{{\em Phys. Rev.
  Lett.} {\bfseries  121} no.~8, (2018) 081301},
  \href{http://arxiv.org/abs/1712.07133}{{\ttfamily arXiv:1712.07133
  [astro-ph.CO]}}.

\bibitem{Nadler:2019zrb}
E.~O. Nadler, V.~Gluscevic, K.~K. Boddy, and R.~H. Wechsler, ``{Constraints on
  Dark Matter Microphysics from the Milky Way Satellite Population},''
  \href{http://dx.doi.org/10.3847/2041-8213/ab1eb2}{{\em Astrophys. J. Lett.}
  {\bfseries  878} no.~2, (2019) 32},
  \href{http://arxiv.org/abs/1904.10000}{{\ttfamily arXiv:1904.10000
  [astro-ph.CO]}}. [Erratum: Astrophys.J.Lett. 897, L46 (2020), Erratum:
  Astrophys.J. 897, L46 (2020)].

\bibitem{Wadekar:2019xnf}
D.~Wadekar and G.~R. Farrar, ``{First astrophysical constraints on dark matter
  interactions with ordinary matter at low relative velocity},''
  \href{http://arxiv.org/abs/1903.12190}{{\ttfamily arXiv:1903.12190
  [hep-ph]}}.

\bibitem{Bhoonah:2018wmw}
A.~Bhoonah, J.~Bramante, F.~Elahi, and S.~Schon, ``{Calorimetric Dark Matter
  Detection With Galactic Center Gas Clouds},''
  \href{http://dx.doi.org/10.1103/PhysRevLett.121.131101}{{\em Phys. Rev.
  Lett.} {\bfseries  121} no.~13, (2018) 131101},
  \href{http://arxiv.org/abs/1806.06857}{{\ttfamily arXiv:1806.06857
  [hep-ph]}}.

\bibitem{Cappiello:2018hsu}
C.~V. Cappiello, K.~C. Ng, and J.~F. Beacom, ``{Reverse Direct Detection:
  Cosmic Ray Scattering With Light Dark Matter},''
  \href{http://dx.doi.org/10.1103/PhysRevD.99.063004}{{\em Phys. Rev. D}
  {\bfseries  99} no.~6, (2019) 063004},
  \href{http://arxiv.org/abs/1810.07705}{{\ttfamily arXiv:1810.07705
  [hep-ph]}}.

\bibitem{Cyburt:2002uw}
R.~H. Cyburt, B.~D. Fields, V.~Pavlidou, and B.~D. Wandelt, ``{Constraining
  strong baryon dark matter interactions with primordial nucleosynthesis and
  cosmic rays},'' \href{http://dx.doi.org/10.1103/PhysRevD.65.123503}{{\em
  Phys. Rev. D} {\bfseries  65} (2002) 123503},
  \href{http://arxiv.org/abs/astro-ph/0203240}{{\ttfamily
  arXiv:astro-ph/0203240}}.

\bibitem{Hooper:2018bfw}
D.~Hooper and S.~D. McDermott, ``{Robust Constraints and Novel Gamma-Ray
  Signatures of Dark Matter That Interacts Strongly With Nucleons},''
  \href{http://dx.doi.org/10.1103/PhysRevD.97.115006}{{\em Phys. Rev. D}
  {\bfseries  97} no.~11, (2018) 115006},
  \href{http://arxiv.org/abs/1802.03025}{{\ttfamily arXiv:1802.03025
  [hep-ph]}}.

\bibitem{Bringmann:2018cvk}
T.~Bringmann and M.~Pospelov, ``{Novel direct detection constraints on light
  dark matter},'' \href{http://dx.doi.org/10.1103/PhysRevLett.122.171801}{{\em
  Phys. Rev. Lett.} {\bfseries  122} no.~17, (2019) 171801},
  \href{http://arxiv.org/abs/1810.10543}{{\ttfamily arXiv:1810.10543
  [hep-ph]}}.

\bibitem{Ema:2018bih}
Y.~Ema, F.~Sala, and R.~Sato, ``{Light Dark Matter at Neutrino Experiments},''
  \href{http://dx.doi.org/10.1103/PhysRevLett.122.181802}{{\em Phys. Rev.
  Lett.} {\bfseries  122} no.~18, (2019) 181802},
  \href{http://arxiv.org/abs/1811.00520}{{\ttfamily arXiv:1811.00520
  [hep-ph]}}.

\bibitem{Cappiello:2019qsw}
C.~Cappiello and J.~F. Beacom, ``{Strong New Limits on Light Dark Matter from
  Neutrino Experiments},''
  \href{http://dx.doi.org/10.1103/PhysRevD.100.103011}{{\em Phys. Rev. D}
  {\bfseries  100} no.~10, (2019) 103011},
  \href{http://arxiv.org/abs/1906.11283}{{\ttfamily arXiv:1906.11283
  [hep-ph]}}.

\bibitem{Dent:2019krz}
J.~B. Dent, B.~Dutta, J.~L. Newstead, and I.~M. Shoemaker, ``{Bounds on Cosmic
  Ray-Boosted Dark Matter in Simplified Models and its Corresponding
  Neutrino-Floor},'' \href{http://dx.doi.org/10.1103/PhysRevD.101.116007}{{\em
  Phys. Rev. D} {\bfseries  101} no.~11, (2020) 116007},
  \href{http://arxiv.org/abs/1907.03782}{{\ttfamily arXiv:1907.03782
  [hep-ph]}}.

\bibitem{Bondarenko:2019vrb}
K.~Bondarenko, A.~Boyarsky, T.~Bringmann, M.~Hufnagel, K.~Schmidt-Hoberg, and
  A.~Sokolenko, ``{Direct detection and complementary constraints for sub-GeV
  dark matter},'' \href{http://dx.doi.org/10.1007/JHEP03(2020)118}{{\em JHEP}
  {\bfseries  03} (2020) 118},
  \href{http://arxiv.org/abs/1909.08632}{{\ttfamily arXiv:1909.08632
  [hep-ph]}}.

\bibitem{Ge:2020yuf}
S.-F. Ge, J.-L. Liu, Q.~Yuan, and N.~Zhou, ``{Boosted Diurnal Effect of Sub-GeV
  Dark Matter at Direct Detection Experiment},''
  \href{http://arxiv.org/abs/2005.09480}{{\ttfamily arXiv:2005.09480
  [hep-ph]}}.

\bibitem{Zhang:2020htl}
B.-L. Zhang, Z.-H. Lei, and J.~Tang, ``{Constraints on cosmic-ray boosted DM in
  CDEX-10},'' \href{http://arxiv.org/abs/2008.07116}{{\ttfamily
  arXiv:2008.07116 [hep-ph]}}.

\bibitem{Wang:2019jtk}
W.~Wang, L.~Wu, J.~M. Yang, H.~Zhou, and B.~Zhu, ``{Sub-GeV Gravity-mediated
  Dark Matter in Direct Detections},''
  \href{http://arxiv.org/abs/1912.09904}{{\ttfamily arXiv:1912.09904
  [hep-ph]}}.

\bibitem{DellaTorre:2016jjf}
S.~Della~Torre {\em et~al.}, ``{From Observations near the Earth to the Local
  Interstellar Spectra},'' in {\em {25th European Cosmic Ray Symposium}}.
\newblock 12, 2016.
\newblock \href{http://arxiv.org/abs/1701.02363}{{\ttfamily arXiv:1701.02363
  [astro-ph.HE]}}.

\bibitem{Bisschoff:2019lne}
D.~Bisschoff, M.~Potgieter, and O.~Aslam, ``{New very local interstellar
  spectra for electrons, positrons, protons and light cosmic ray nuclei},''
  \href{http://dx.doi.org/10.3847/1538-4357/ab1e4a}{{\em Astrophys. J.}
  {\bfseries  878} no.~1, (2019) 59},
  \href{http://arxiv.org/abs/1902.10438}{{\ttfamily arXiv:1902.10438
  [astro-ph.HE]}}.

\bibitem{Strong:1998pw}
A.~Strong and I.~Moskalenko, ``{Propagation of cosmic-ray nucleons in the
  galaxy},'' \href{http://dx.doi.org/10.1086/306470}{{\em Astrophys. J.}
  {\bfseries  509} (1998) 212--228},
  \href{http://arxiv.org/abs/astro-ph/9807150}{{\ttfamily
  arXiv:astro-ph/9807150}}.

\bibitem{Moskalenko:1997gh}
I.~Moskalenko and A.~Strong, ``{Production and propagation of cosmic ray
  positrons and electrons},'' \href{http://dx.doi.org/10.1086/305152}{{\em
  Astrophys. J.} }, \href{http://arxiv.org/abs/astro-ph/9710124}{{\ttfamily
  arXiv:astro-ph/9710124}}.

\bibitem{Adriani:2011cu}
{\bfseries  PAMELA} Collaboration, O.~Adriani {\em et~al.}, ``{PAMELA
  Measurements of Cosmic-ray Proton and Helium Spectra},''
  \href{http://dx.doi.org/10.1126/science.1199172}{{\em Science} {\bfseries
  332} (2011) 69--72}, \href{http://arxiv.org/abs/1103.4055}{{\ttfamily
  arXiv:1103.4055 [astro-ph.HE]}}.

\bibitem{Adriani:2013as}
O.~Adriani {\em et~al.}, ``{Time dependence of the proton flux measured by
  PAMELA during the July 2006 - December 2009 solar minimum},''
  \href{http://dx.doi.org/10.1088/0004-637X/765/2/91}{{\em Astrophys. J.}
  {\bfseries  765} (2013) 91}, \href{http://arxiv.org/abs/1301.4108}{{\ttfamily
  arXiv:1301.4108 [astro-ph.HE]}}.

\bibitem{Aguilar:2015ooa}
{\bfseries  AMS} Collaboration, M.~Aguilar {\em et~al.}, ``{Precision
  Measurement of the Proton Flux in Primary Cosmic Rays from Rigidity 1 GV to
  1.8 TV with the Alpha Magnetic Spectrometer on the International Space
  Station},'' \href{http://dx.doi.org/10.1103/PhysRevLett.114.171103}{{\em
  Phys. Rev. Lett.} {\bfseries  114} (2015) 171103}.

\bibitem{Aguilar:2015ctt}
{\bfseries  AMS} Collaboration, M.~Aguilar {\em et~al.}, ``{Precision
  Measurement of the Helium Flux in Primary Cosmic Rays of Rigidities 1.9 GV to
  3 TV with the Alpha Magnetic Spectrometer on the International Space
  Station},'' \href{http://dx.doi.org/10.1103/PhysRevLett.115.211101}{{\em
  Phys. Rev. Lett.} {\bfseries  115} no.~21, (2015) 211101}.

\bibitem{Yoon:2011aa}
Y.~Yoon {\em et~al.}, ``{Cosmic-Ray Proton and Helium Spectra from the First
  CREAM Flight},'' \href{http://dx.doi.org/10.1088/0004-637X/728/2/122}{{\em
  Astrophys. J.} {\bfseries  728} (2011) 122},
  \href{http://arxiv.org/abs/1102.2575}{{\ttfamily arXiv:1102.2575
  [astro-ph.HE]}}.

\bibitem{Yoon:2017qjx}
Y.~Yoon {\em et~al.}, ``{Proton and Helium Spectra from the CREAM-III
  Flight},'' \href{http://dx.doi.org/10.3847/1538-4357/aa68e4}{{\em Astrophys.
  J.} {\bfseries  839} no.~1, (2017) 5},
  \href{http://arxiv.org/abs/1704.02512}{{\ttfamily arXiv:1704.02512
  [astro-ph.HE]}}.

\bibitem{Adriani:2019aft}
{\bfseries  CALET} Collaboration, O.~Adriani {\em et~al.}, ``{Direct
  Measurement of the Cosmic-Ray Proton Spectrum from 50 GeV to 10 TeV with the
  Calorimetric Electron Telescope on the International Space Station},''
  \href{http://dx.doi.org/10.1103/PhysRevLett.122.181102}{{\em Phys. Rev.
  Lett.} {\bfseries  122} no.~18, (2019) 181102},
  \href{http://arxiv.org/abs/1905.04229}{{\ttfamily arXiv:1905.04229
  [astro-ph.HE]}}.

\bibitem{An:2019wcw}
{\bfseries  DAMPE} Collaboration, Q.~An {\em et~al.}, ``{Measurement of the
  cosmic-ray proton spectrum from 40 GeV to 100 TeV with the DAMPE
  satellite},'' \href{http://dx.doi.org/10.1126/sciadv.aax3793}{{\em Sci. Adv.}
  {\bfseries  5} no.~9, (2019) eaax3793},
  \href{http://arxiv.org/abs/1909.12860}{{\ttfamily arXiv:1909.12860
  [astro-ph.HE]}}.

\bibitem{Nilsen:1997mv}
B.~Nilsen {\em et~al.}, ``{Cosmic ray H and He spectra from 2-TeV/nucleon to
  800-TeV/nucleon from the JACEE experiments},''
  \href{http://dx.doi.org/10.1063/1.54396}{{\em AIP Conf. Proc.} {\bfseries
  412} no.~1, (1997) 1031--1034}.

\bibitem{Derbina:2005ta}
{\bfseries  RUNJOB} Collaboration, V.~Derbina {\em et~al.}, ``{Cosmic-ray
  spectra and composition in the energy range of 10-TeV - 1000-TeV per particle
  obtained by the RUNJOB experiment},''
  \href{http://dx.doi.org/10.1086/432715}{{\em Astrophys. J. Lett.} {\bfseries
  628} (2005) L41--L44}.

\bibitem{Atkin:2018wsp}
E.~Atkin {\em et~al.}, ``{New Universal Cosmic-Ray Knee near a Magnetic
  Rigidity of 10 TV with the NUCLEON Space Observatory},''
  \href{http://dx.doi.org/10.1134/S0021364018130015}{{\em JETP Lett.}
  {\bfseries  108} no.~1, (2018) 5--12},
  \href{http://arxiv.org/abs/1805.07119}{{\ttfamily arXiv:1805.07119
  [astro-ph.HE]}}.

\bibitem{Greisen:1966jv}
K.~Greisen, ``{End to the cosmic ray spectrum?},''
  \href{http://dx.doi.org/10.1103/PhysRevLett.16.748}{{\em Phys. Rev. Lett.}
  {\bfseries  16} (1966) 748--750}.

\bibitem{Zatsepin:1966xyz}
G. T. Zatsepin and V. A. Kuzmin, JETP Lett. 4, 78 (1966).

\bibitem{Aharonian:2009ah}
{\bfseries  H.E.S.S.} Collaboration, F.~Aharonian {\em et~al.}, ``{Probing the
  ATIC peak in the cosmic-ray electron spectrum with H.E.S.S},''
  \href{http://dx.doi.org/10.1051/0004-6361/200913323}{{\em Astron. Astrophys.}
  {\bfseries  508} (2009) 561},
  \href{http://arxiv.org/abs/0905.0105}{{\ttfamily arXiv:0905.0105
  [astro-ph.HE]}}.

\bibitem{Adriani:2017efm}
{\bfseries  CALET} Collaboration, O.~Adriani {\em et~al.}, ``{Energy Spectrum
  of Cosmic-Ray Electron and Positron from 10 GeV to 3 TeV Observed with the
  Calorimetric Electron Telescope on the International Space Station},''
  \href{http://dx.doi.org/10.1103/PhysRevLett.119.181101}{{\em Phys. Rev.
  Lett.} {\bfseries  119} no.~18, (2017) 181101},
  \href{http://arxiv.org/abs/1712.01711}{{\ttfamily arXiv:1712.01711
  [astro-ph.HE]}}.

\bibitem{Ambrosi:2017wek}
{\bfseries  DAMPE} Collaboration, G.~Ambrosi {\em et~al.}, ``{Direct detection
  of a break in the teraelectronvolt cosmic-ray spectrum of electrons and
  positrons},'' \href{http://dx.doi.org/10.1038/nature24475}{{\em Nature}
  {\bfseries  552} (2017) 63--66},
  \href{http://arxiv.org/abs/1711.10981}{{\ttfamily arXiv:1711.10981
  [astro-ph.HE]}}.

\bibitem{Abbasi:2007sv}
{\bfseries  HiRes} Collaboration, R.~Abbasi {\em et~al.}, ``{First observation
  of the Greisen-Zatsepin-Kuzmin suppression},''
  \href{http://dx.doi.org/10.1103/PhysRevLett.100.101101}{{\em Phys. Rev.
  Lett.} {\bfseries  100} (2008) 101101},
  \href{http://arxiv.org/abs/astro-ph/0703099}{{\ttfamily
  arXiv:astro-ph/0703099}}.

\bibitem{Castellina:2019huz}
{\bfseries  Pierre Auger} Collaboration, A.~Castellina, ``{Highlights from the
  Pierre Auger Observatory (ICRC2019)},''
  \href{http://dx.doi.org/10.22323/1.358.0004}{{\em PoS} {\bfseries  ICRC2019}
  (2020) 004}, \href{http://arxiv.org/abs/1909.10791}{{\ttfamily
  arXiv:1909.10791 [astro-ph.HE]}}.

\bibitem{Deligny:2020gzq}
{\bfseries  Pierre Auger, Telescope Array} Collaboration, O.~Deligny, ``{The
  energy spectrum of ultra-high energy cosmic rays measured at the Pierre Auger
  Observatory and at the Telescope Array},''
  \href{http://dx.doi.org/10.22323/1.358.0234}{{\em PoS} {\bfseries  ICRC2019}
  (2020) 234}, \href{http://arxiv.org/abs/2001.08811}{{\ttfamily
  arXiv:2001.08811 [astro-ph.HE]}}.

\bibitem{Kachelriess:2019oqu}
M.~Kachelriess and D.~Semikoz, ``{Cosmic Ray Models},''
  \href{http://dx.doi.org/10.1016/j.ppnp.2019.07.002}{{\em Prog. Part. Nucl.
  Phys.} {\bfseries  109} (2019) 103710},
  \href{http://arxiv.org/abs/1904.08160}{{\ttfamily arXiv:1904.08160
  [astro-ph.HE]}}.

\bibitem{Watson:2013cla}
A.~A. Watson, ``{High-energy cosmic rays and the Greisen--Zatsepin--Kuz'min
  effect},'' \href{http://dx.doi.org/10.1088/0034-4885/77/3/036901}{{\em Rept.
  Prog. Phys.} {\bfseries  77} (2014) 036901},
  \href{http://arxiv.org/abs/1310.0325}{{\ttfamily arXiv:1310.0325
  [astro-ph.HE]}}.

\bibitem{Nagano:2000ve}
M.~Nagano and A.~A. Watson, ``{Observations and implications of the
  ultrahigh-energy cosmic rays},''
  \href{http://dx.doi.org/10.1103/RevModPhys.72.689}{{\em Rev. Mod. Phys.}
  {\bfseries  72} (2000) 689--732}.

\bibitem{Gaisser:2013bla}
T.~K. Gaisser, T.~Stanev, and S.~Tilav, ``{Cosmic Ray Energy Spectrum from
  Measurements of Air Showers},''
  \href{http://dx.doi.org/10.1007/s11467-013-0319-7}{{\em Front. Phys.
  (Beijing)} {\bfseries  8} (2013) 748--758},
  \href{http://arxiv.org/abs/1303.3565}{{\ttfamily arXiv:1303.3565
  [astro-ph.HE]}}.

\bibitem{Peters:1966xyz}
Peters, Il Nuovo Cim. XXII, 800-819 (1961).

\bibitem{Nagano:1991jz}
M.~Nagano, M.~Teshima, Y.~Matsubara, H.~Dai, T.~Hara, N.~Hayashida, M.~Honda,
  H.~Ohoka, and S.~Yoshida, ``{Energy spectrum of primary cosmic rays above
  10**17-eV determined from the extensive air shower experiment at Akeno},''
  \href{http://dx.doi.org/10.1088/0954-3899/18/2/022}{{\em J. Phys. G}
  {\bfseries  18} (1992) 423--442}.

\bibitem{Glasmacher:1999id}
M.~Glasmacher {\em et~al.}, ``{The cosmic ray energy spectrum between 10**14-eV
  and 10**16-eV},'' \href{http://dx.doi.org/10.1016/S0927-6505(98)00070-X}{{\em
  Astropart. Phys.} {\bfseries  10} (1999) 291--302}.

\bibitem{Arqueros:1999uq}
{\bfseries  HEGRA} Collaboration, F.~Arqueros {\em et~al.}, ``{Energy spectrum
  and chemical composition of cosmic rays between 0.3-PeV and 10-PeV determined
  from the Cherenkov light and charged particle distributions in air
  showers},'' {\em Astron. Astrophys.} {\bfseries  359} (2000) 682--694,
  \href{http://arxiv.org/abs/astro-ph/9908202}{{\ttfamily
  arXiv:astro-ph/9908202}}.

\bibitem{Takeda:2002at}
M.~Takeda {\em et~al.}, ``{Energy determination in the Akeno Giant Air Shower
  Array experiment},''
  \href{http://dx.doi.org/10.1016/S0927-6505(02)00243-8}{{\em Astropart. Phys.}
  {\bfseries  19} (2003) 447--462},
  \href{http://arxiv.org/abs/astro-ph/0209422}{{\ttfamily
  arXiv:astro-ph/0209422}}.

\bibitem{Antoni:2005wq}
{\bfseries  KASCADE} Collaboration, T.~Antoni {\em et~al.}, ``{KASCADE
  measurements of energy spectra for elemental groups of cosmic rays: Results
  and open problems},''
  \href{http://dx.doi.org/10.1016/j.astropartphys.2005.04.001}{{\em Astropart.
  Phys.} {\bfseries  24} (2005) 1--25},
  \href{http://arxiv.org/abs/astro-ph/0505413}{{\ttfamily
  arXiv:astro-ph/0505413}}.

\bibitem{Amenomori:2008aa}
{\bfseries  TIBET III} Collaboration, M.~Amenomori {\em et~al.}, ``{The
  All-particle spectrum of primary cosmic rays in the wide energy range from
  10**14 eV to 10**17 eV observed with the Tibet-III air-shower array},''
  \href{http://dx.doi.org/10.1086/529514}{{\em Astrophys. J.} {\bfseries  678}
  (2008) 1165--1179}, \href{http://arxiv.org/abs/0801.1803}{{\ttfamily
  arXiv:0801.1803 [hep-ex]}}.

\bibitem{Aartsen:2013wda}
{\bfseries  IceCube} Collaboration, M.~Aartsen {\em et~al.}, ``{Measurement of
  the cosmic ray energy spectrum with IceTop-73},''
  \href{http://dx.doi.org/10.1103/PhysRevD.88.042004}{{\em Phys. Rev. D}
  {\bfseries  88} no.~4, (2013) 042004},
  \href{http://arxiv.org/abs/1307.3795}{{\ttfamily arXiv:1307.3795
  [astro-ph.HE]}}.

\bibitem{Ter-Antonyan:2014hea}
S.~Ter-Antonyan, ``{Sharp knee phenomenon of primary cosmic ray energy
  spectrum},'' \href{http://dx.doi.org/10.1103/PhysRevD.89.123003}{{\em Phys.
  Rev. D} {\bfseries  89} no.~12, (2014) 123003},
  \href{http://arxiv.org/abs/1405.5472}{{\ttfamily arXiv:1405.5472
  [astro-ph.HE]}}.

\bibitem{Prosin:2016rqu}
V.~Prosin {\em et~al.}, ``{Results from Tunka-133 (5 years observation) and
  from the Tunka-HiSCORE prototype},''
  \href{http://dx.doi.org/10.1051/epjconf/201612103004}{{\em EPJ Web Conf.}
  {\bfseries  121} (2016) 03004}.

\bibitem{Fenu:2017hlc}
{\bfseries  Pierre Auger} Collaboration, F.~Fenu, ``{The cosmic ray energy
  spectrum measured using the Pierre Auger Observatory},''
  \href{http://dx.doi.org/10.22323/1.301.0486}{{\em PoS} {\bfseries  ICRC2017}
  (2018) 486}.

\bibitem{Arteaga-Velazquez:2017rmh}
{\bfseries  KASCADE Grande} Collaboration, C.~J. Arteaga-Velázquez {\em
  et~al.}, ``{Measurements of the muon content of EAS in KASCADE-Grande
  compared with SIBYLL 2.3 predictions},''
  \href{http://dx.doi.org/10.22323/1.301.0316}{{\em PoS} {\bfseries  ICRC2017}
  (2018) 316}.

\bibitem{Perdrisat:2006hj}
C.~Perdrisat, V.~Punjabi, and M.~Vanderhaeghen, ``{Nucleon Electromagnetic Form
  Factors},'' \href{http://dx.doi.org/10.1016/j.ppnp.2007.05.001}{{\em Prog.
  Part. Nucl. Phys.} {\bfseries  59} (2007) 694--764},
  \href{http://arxiv.org/abs/hep-ph/0612014}{{\ttfamily arXiv:hep-ph/0612014}}.

\bibitem{Helm:1956zz}
R.~H. Helm, ``{Inelastic and Elastic Scattering of 187-Mev Electrons from
  Selected Even-Even Nuclei},''
  \href{http://dx.doi.org/10.1103/PhysRev.104.1466}{{\em Phys. Rev.} {\bfseries
   104} (1956) 1466--1475}.

\bibitem{Lewin:1995rx}
J.~Lewin and P.~Smith, ``{Review of mathematics, numerical factors, and
  corrections for dark matter experiments based on elastic nuclear recoil},''
  \href{http://dx.doi.org/10.1016/S0927-6505(96)00047-3}{{\em Astropart. Phys.}
  {\bfseries  6} (1996) 87--112}.

\bibitem{Kouvaris:2014lpa}
C.~Kouvaris and I.~M. Shoemaker, ``{Daily modulation as a smoking gun of dark
  matter with significant stopping rate},''
  \href{http://dx.doi.org/10.1103/PhysRevD.90.095011}{{\em Phys. Rev. D}
  {\bfseries  90} (2014) 095011},
  \href{http://arxiv.org/abs/1405.1729}{{\ttfamily arXiv:1405.1729 [hep-ph]}}.

\bibitem{Starkman:1990nj}
G.~D. Starkman, A.~Gould, R.~Esmailzadeh, and S.~Dimopoulos, ``{Opening the
  Window on Strongly Interacting Dark Matter},''
  \href{http://dx.doi.org/10.1103/PhysRevD.41.3594}{{\em Phys. Rev. D}
  {\bfseries  41} (1990) 3594}.

\bibitem{Rudnick:2003xyz}
R. Rudnick and S. Gao, Composition of the continental crust, in Treatise on
  Geochemistry, Pergamon, Oxford, 2003.DOI:10.1016/B0-08-043751-6/03016-4.

\bibitem{Aprile:2018dbl}
{\bfseries  XENON} Collaboration, E.~Aprile {\em et~al.}, ``{Dark Matter Search
  Results from a One Ton-Year Exposure of XENON1T},''
  \href{http://dx.doi.org/10.1103/PhysRevLett.121.111302}{{\em Phys. Rev.
  Lett.} {\bfseries  121} no.~11, (2018) 111302},
  \href{http://arxiv.org/abs/1805.12562}{{\ttfamily arXiv:1805.12562
  [astro-ph.CO]}}.

\bibitem{Aprile:2019xxb}
{\bfseries  XENON} Collaboration, E.~Aprile {\em et~al.}, ``{Light Dark Matter
  Search with Ionization Signals in XENON1T},''
  \href{http://dx.doi.org/10.1103/PhysRevLett.123.251801}{{\em Phys. Rev.
  Lett.} {\bfseries  123} no.~25, (2019) 251801},
  \href{http://arxiv.org/abs/1907.11485}{{\ttfamily arXiv:1907.11485
  [hep-ex]}}.

\bibitem{Aprile:2019dme}
{\bfseries  XENON} Collaboration, E.~Aprile {\em et~al.}, ``{XENON1T dark
  matter data analysis: Signal and background models and statistical
  inference},'' \href{http://dx.doi.org/10.1103/PhysRevD.99.112009}{{\em Phys.
  Rev. D} {\bfseries  99} no.~11, (2019) 112009},
  \href{http://arxiv.org/abs/1902.11297}{{\ttfamily arXiv:1902.11297
  [physics.ins-det]}}.

\bibitem{Green:2001xy}
A.~M. Green, ``{Calculating exclusion limits for weakly interacting massive
  particle direct detection experiments without background subtraction},''
  \href{http://dx.doi.org/10.1103/PhysRevD.65.023520}{{\em Phys. Rev. D}
  {\bfseries  65} (2002) 023520},
  \href{http://arxiv.org/abs/astro-ph/0106555}{{\ttfamily
  arXiv:astro-ph/0106555}}.

\bibitem{Savage:2008er}
C.~Savage, G.~Gelmini, P.~Gondolo, and K.~Freese, ``{Compatibility of
  DAMA/LIBRA dark matter detection with other searches},''
  \href{http://dx.doi.org/10.1088/1475-7516/2009/04/010}{{\em JCAP} {\bfseries
  04} (2009) 010}, \href{http://arxiv.org/abs/0808.3607}{{\ttfamily
  arXiv:0808.3607 [astro-ph]}}.

\bibitem{Angloher:2015ewa}
{\bfseries  CRESST} Collaboration, G.~Angloher {\em et~al.}, ``{Results on
  light dark matter particles with a low-threshold CRESST-II detector},''
  \href{http://dx.doi.org/10.1140/epjc/s10052-016-3877-3}{{\em Eur. Phys. J. C}
  {\bfseries  76} no.~1, (2016) 25},
  \href{http://arxiv.org/abs/1509.01515}{{\ttfamily arXiv:1509.01515
  [astro-ph.CO]}}.

\bibitem{Angloher:2017sxg}
{\bfseries  CRESST} Collaboration, G.~Angloher {\em et~al.}, ``{Results on
  MeV-scale dark matter from a gram-scale cryogenic calorimeter operated above
  ground},'' \href{http://dx.doi.org/10.1140/epjc/s10052-017-5223-9}{{\em Eur.
  Phys. J. C} {\bfseries  77} no.~9, (2017) 637},
  \href{http://arxiv.org/abs/1707.06749}{{\ttfamily arXiv:1707.06749
  [astro-ph.CO]}}.

\bibitem{Mahdawi:2018euy}
M.~S. Mahdawi and G.~R. Farrar, ``{Constraints on Dark Matter with a moderately
  large and velocity-dependent DM-nucleon cross-section},''
  \href{http://dx.doi.org/10.1088/1475-7516/2018/10/007}{{\em JCAP} {\bfseries
  10} (2018) 007}, \href{http://arxiv.org/abs/1804.03073}{{\ttfamily
  arXiv:1804.03073 [hep-ph]}}.

\bibitem{Xu:2018efh}
W.~L. Xu, C.~Dvorkin, and A.~Chael, ``{Probing sub-GeV Dark Matter-Baryon
  Scattering with Cosmological Observables},''
  \href{http://dx.doi.org/10.1103/PhysRevD.97.103530}{{\em Phys. Rev. D}
  {\bfseries  97} no.~10, (2018) 103530},
  \href{http://arxiv.org/abs/1802.06788}{{\ttfamily arXiv:1802.06788
  [astro-ph.CO]}}.

\bibitem{Aguilar:2017hno}
{\bfseries  AMS} Collaboration, M.~Aguilar {\em et~al.}, ``{Observation of the
  Identical Rigidity Dependence of He, C, and O Cosmic Rays at High Rigidities
  by the Alpha Magnetic Spectrometer on the International Space Station},''
  \href{http://dx.doi.org/10.1103/PhysRevLett.119.251101}{{\em Phys. Rev.
  Lett.} {\bfseries  119} no.~25, (2017) 251101}.

\bibitem{Slatyer:2018aqg}
T.~R. Slatyer and C.-L. Wu, ``{Early-Universe constraints on dark matter-baryon
  scattering and their implications for a global 21 cm signal},''
  \href{http://dx.doi.org/10.1103/PhysRevD.98.023013}{{\em Phys. Rev. D}
  {\bfseries  98} no.~2, (2018) 023013},
  \href{http://arxiv.org/abs/1803.09734}{{\ttfamily arXiv:1803.09734
  [astro-ph.CO]}}.

\bibitem{Boddy:2018kfv}
K.~K. Boddy and V.~Gluscevic, ``{First Cosmological Constraint on the Effective
  Theory of Dark Matter-Proton Interactions},''
  \href{http://dx.doi.org/10.1103/PhysRevD.98.083510}{{\em Phys. Rev. D}
  {\bfseries  98} no.~8, (2018) 083510},
  \href{http://arxiv.org/abs/1801.08609}{{\ttfamily arXiv:1801.08609
  [astro-ph.CO]}}.

\bibitem{Murgia:2018now}
R.~Murgia, V.~Ir\v~si\v c, and M.~Viel, ``{Novel constraints on noncold,
  nonthermal dark matter from Lyman- $\alpha$ forest data},''
  \href{http://dx.doi.org/10.1103/PhysRevD.98.083540}{{\em Phys. Rev. D}
  {\bfseries  98} no.~8, (2018) 083540},
  \href{http://arxiv.org/abs/1806.08371}{{\ttfamily arXiv:1806.08371
  [astro-ph.CO]}}.

\bibitem{Krnjaic:2019dzc}
G.~Krnjaic and S.~D. McDermott, ``{Implications of BBN Bounds for Cosmic Ray
  Upscattered Dark Matter},''
  \href{http://dx.doi.org/10.1103/PhysRevD.101.123022}{{\em Phys. Rev. D}
  {\bfseries  101} no.~12, (2020) 123022},
  \href{http://arxiv.org/abs/1908.00007}{{\ttfamily arXiv:1908.00007
  [hep-ph]}}.

\bibitem{Ahn:2010gv}
{\bfseries  CREAM} Collaboration, H.~Ahn {\em et~al.}, ``{Discrepant hardening
  observed in cosmic-ray elemental spectra},''
  \href{http://dx.doi.org/10.1088/2041-8205/714/1/L89}{{\em Astrophys. J.
  Lett.} {\bfseries  714} (2010) L89--L93},
  \href{http://arxiv.org/abs/1004.1123}{{\ttfamily arXiv:1004.1123
  [astro-ph.HE]}}.

\bibitem{Gaisser:2012zz}
T.~K. Gaisser, ``{Spectrum of cosmic-ray nucleons, kaon production, and the
  atmospheric muon charge ratio},''
  \href{http://dx.doi.org/10.1016/j.astropartphys.2012.02.010}{{\em Astropart.
  Phys.} {\bfseries  35} (2012) 801--806},
  \href{http://arxiv.org/abs/1111.6675}{{\ttfamily arXiv:1111.6675
  [astro-ph.HE]}}.

\bibitem{Hillas:2005cs}
A.~Hillas, ``{Can diffusive shock acceleration in supernova remnants account
  for high-energy galactic cosmic rays?},''
  \href{http://dx.doi.org/10.1088/0954-3899/31/5/R02}{{\em J. Phys. G}
  {\bfseries  31} (2005) R95--R131}.

\bibitem{Lenardo:2014cva}
B.~Lenardo, K.~Kazkaz, A.~Manalaysay, J.~Mock, M.~Szydagis, and M.~Tripathi,
  ``{A Global Analysis of Light and Charge Yields in Liquid Xenon},''
  \href{http://dx.doi.org/10.1109/TNS.2015.2481322}{{\em IEEE Trans. Nucl.
  Sci.} {\bfseries  62} no.~6, (2015) 3387--3396},
  \href{http://arxiv.org/abs/1412.4417}{{\ttfamily arXiv:1412.4417
  [astro-ph.IM]}}.

\bibitem{Thomas:1987zz}
J.~Thomas and D.~Imel, ``{Recombination of electron-ion pairs in liquid argon
  and liquid xenon},'' \href{http://dx.doi.org/10.1103/PhysRevA.36.614}{{\em
  Phys. Rev. A} {\bfseries  36} (1987) 614--616}.

\bibitem{Szydagis:2011tk}
M.~Szydagis, N.~Barry, K.~Kazkaz, J.~Mock, D.~Stolp, M.~Sweany, M.~Tripathi,
  S.~Uvarov, N.~Walsh, and M.~Woods, ``{NEST: A Comprehensive Model for
  Scintillation Yield in Liquid Xenon},''
  \href{http://dx.doi.org/10.1088/1748-0221/6/10/P10002}{{\em JINST} {\bfseries
   6} (2011) P10002}, \href{http://arxiv.org/abs/1106.1613}{{\ttfamily
  arXiv:1106.1613 [physics.ins-det]}}.

\bibitem{Drukier:1986tm}
A.~Drukier, K.~Freese, and D.~Spergel, ``{Detecting Cold Dark Matter
  Candidates},'' \href{http://dx.doi.org/10.1103/PhysRevD.33.3495}{{\em Phys.
  Rev. D} {\bfseries  33} (1986) 3495--3508}.

\bibitem{Rolke:2004mj}
W.~A. Rolke, A.~M. Lopez, and J.~Conrad, ``{Limits and confidence intervals in
  the presence of nuisance parameters},''
  \href{http://dx.doi.org/10.1016/j.nima.2005.05.068}{{\em Nucl. Instrum. Meth.
  A} {\bfseries  551} (2005) 493--503},
  \href{http://arxiv.org/abs/physics/0403059}{{\ttfamily
  arXiv:physics/0403059}}.

\bibitem{Aprile:2019bbb}
{\bfseries  XENON} Collaboration, E.~Aprile {\em et~al.}, ``{XENON1T Dark
  Matter Data Analysis: Signal Reconstruction, Calibration and Event
  Selection},'' \href{http://dx.doi.org/10.1103/PhysRevD.100.052014}{{\em Phys.
  Rev. D} {\bfseries  100} no.~5, (2019) 052014},
  \href{http://arxiv.org/abs/1906.04717}{{\ttfamily arXiv:1906.04717
  [physics.ins-det]}}.

\end{thebibliography}\endgroup

\end{document}